\documentclass{goose-article}

\hypersetup{pdfauthor={T.W.J. de Geus}}
\header{%
  T.W.J.~de~Geus, R.H.J.~Peerlings, M.G.D.~Geers\\
  Engineering Fracture Mechanics, 2017, 169:354--370, \doi{10.1016/j.engfracmech.2016.08.009}, \eprint{1603.08910}
}

\title{%
  Fracture in multi-phase materials:\texorpdfstring{\\}{} why some microstructures are more critical than others
}
\author[1,2]{T.W.J.~de~Geus$^*$}
\author[2]{R.H.J.~Peerlings}
\author[2]{M.G.D.~Geers}
\affil[1]{%
  Materials innovation institute (M2i), Delft, The Netherlands%
}
\affil[2]{%
  Department of Mechanical Engineering, Eindhoven University of Technology, Eindhoven, The Netherlands%
}
\contact{%
  $^*$Corresponding author:
  \href{mailto:t.w.j.d.geus@tue.nl}{t.w.j.d.geus@tue.nl} %
  \hspace{1mm}--\hspace{1mm} %
  \href{mailto:tom@geus.me}{tom@geus.me} %
  \hspace{1mm}--\hspace{1mm} %
  \href{http://www.geus.me}{www.geus.me}%
}

\begin{document}

\maketitle

\begin{abstract}
Our goal is to unravel the mechanisms that lead to failure of a ductile two-phase material -- that consists of a ductile soft phase and a relatively brittle hard phase. An idealized microstructural model is used to study damage propagation systematically and transparently. The analysis uncovers distinct microstructural features around early voids, whereby regions of the hard phase are aligned with the tensile axis and regions of the soft phase are aligned with the shear directions. These features are consistently found in regions exhibiting damage propagation, whereby the damage remains initiation driven, i.e.\ voids nucleate independently of each other. Upon localization, damage is controlled on a longer length-scale relying on a critical relative position of `initiation hot-spots'. The damage rapidly increases in bands of the soft phase wherein several voids are aligned with the shear directions. The relative arrangement of the voids determines whether the microstructure fails early, or at a substantially higher strain. Although much research is needed to refine these findings for real or more realistic microstructures, in particular in three-dimensions, this paper opens a route to a deeper understanding of the ultimate failure of multi-phase materials.
\end{abstract}

\keywords{\\
ductile fracture; damage propagation; multi-phase materials; spatial correlation; micromechanics}

\section{Introduction}

It is well established that the evolution of damage in ductile multi-phase materials is strongly influenced by local microstructural features such as the spatial distribution of the phases \citep{Yan2015,LLorca2004,Li1999}. Mechanically speaking these materials comprise one or more hard phase(s) reinforcing one or more soft phase(s). In particular clustering of the hard phase is found to promote void nucleation \cite{Lewandowski1989,Kim1981,Kim2000,Avramovic-Cingara2009,Tasan2010}, while also the more global hard phase volume fraction is found to correlate with the ductility \cite{Davies1978a,Lawson1981,Speich1979,Ahmad2000,Llorca1991,LeRoy1981}.

Voids generally nucleate throughout the microstructure at all stages of deformation, but only some of them rapidly coalesce into a macroscopic crack at the final stage of deformation \citep{LeRoy1981,Lewandowski1989,Avramovic-Cingara2009a,Kadkhodapour2011,Prahl2007}. Hence, only a small fraction of the nucleated voids actually contribute to the final fracture. The question that thus arises is: what governs the ultimate fracture? Which clusters of voids sufficiently weaken the microstructure to trigger localization? Is the initiation of final fracture predetermined by critical microstructural features? How is this influenced by the relative amount and relative mechanical contrast of the phases? This paper aims to answer these questions using a systematic numerical analysis. An artificial microstructure is used in which the two phases are randomly distributed in a regular grid of square cells. Although such microstructures are of course not very realistic, they enable a transparent analysis, in which mechanisms that consistently occur, in a large set of random microstructures, are naturally identified as damage `hot-spots'. The idealized microstructure also allows us to systematically vary microstructural parameters, without any bias or cross-talk due to experimental limitations. Compared to other studies in the damage propagation regime \citep[e.g.][]{Vajragupta2012,Lewandowski1989}, this paper adds a statistical perspective. While compared to earlier statistical studies \citep[e.g.][]{Kumar2006,DeGeus2015a} this paper goes beyond the stage of fracture initiation.

To enable the statistical analysis, the mechanics are modeled in a simple and efficient way. The two phases are treated as isotropic elasto-plastic, and the damage is modeled by a simple indicator. Based on the value of the damage indicator, individual cells are eroded from the microstructure without prior softening. This implies that the cell size is used to regularize the damage. A statistically representative ensemble of random periodic volume elements are considered for which all conditions are identical, and only the microstructural arrangement of the phases differs.

The adopted periodicity does not allow an analysis beyond the loss of stability, in the post-critical regime \citep{LLorca2004}. Remedies for this are available in the literature \citep{Coenen2011,Kouznetsova2002a,Ghosh2009}. However, the added computational complexity render them unattractive for the statistical analysis carried out here. Sufficiently large volume elements are therefore used with conventional periodic boundary conditions and the analysis is restricted to the nucleation and early propagation of defects, up to the onset of instability.

The paper is structured as follows. The model, including damage propagation by cell erosion, is discussed in Section~\ref{sec:model}. In Section~\ref{sec:reference}, the influence of the microstructure, including that of early nucleated voids, on localization is studied for a reference case with constant properties in terms of hard phase volume fraction and phase contrast. Subsequently, these two parameters are varied in Section~\ref{sec:parameter}. Section~\ref{sec:triax} studies the competition between fracture of the soft and of the hard phase, under various applied stress triaxialities. This is followed by a brief discussion and concluding remarks in Sections~\ref{sec:discusion} and \ref{sec:conclusion}.

\section{Model}
\label{sec:model}

\subsection{Microstructure \& discretization}

\begin{figure}[htbp]
  \centering
  \includegraphics[width=0.7\textwidth]{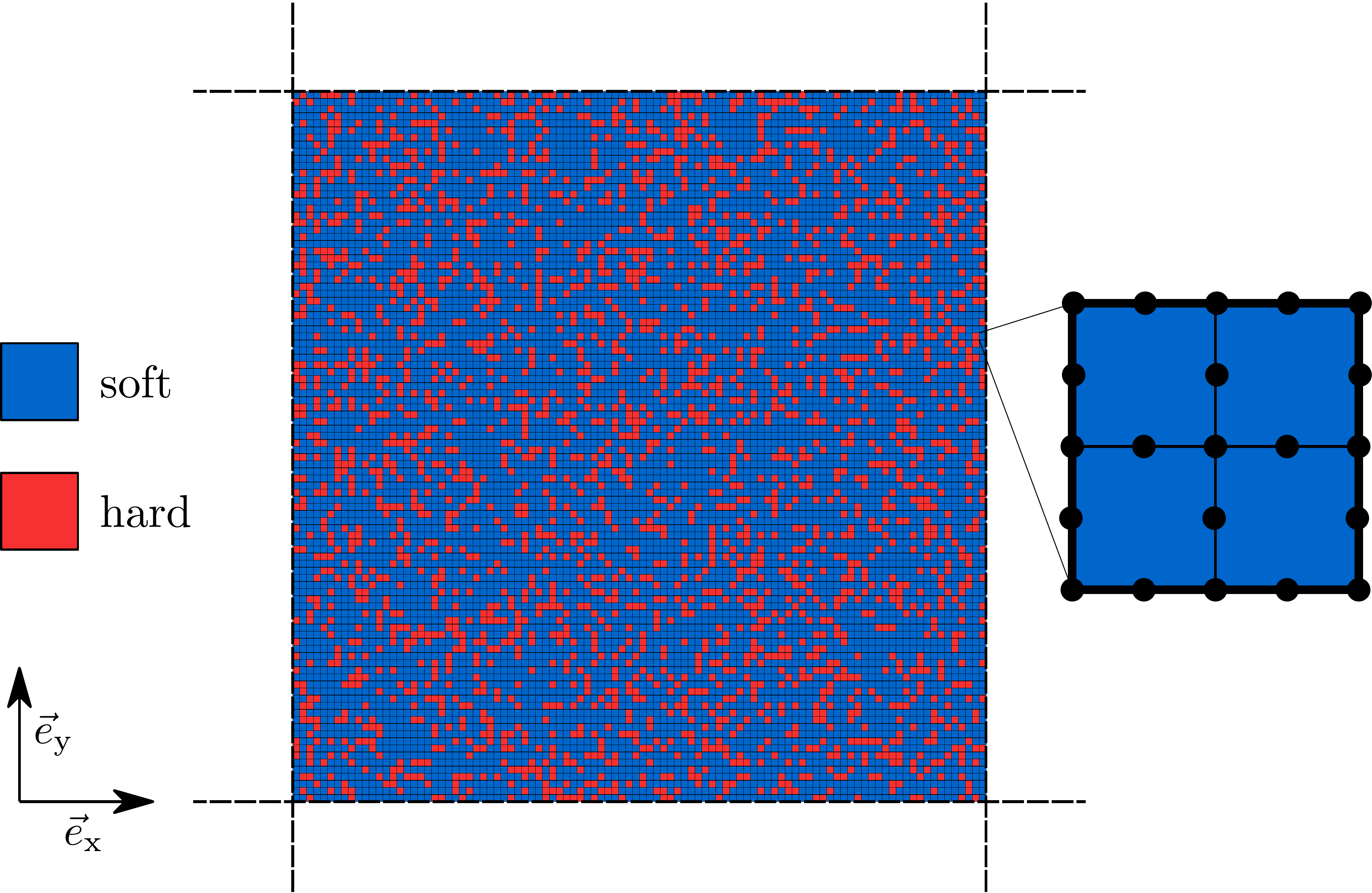}
  \caption{A typical microstructure (periodic volume element) with $N = 100 \times 100$ cells, from the reference ensemble with a hard phase volume fraction $\varphi_\mathrm{hard} = 0.25$. The finite element discretization of one cell is shown on the right. Each finite element is numerically integrated using four Gauss points.}
  \label{fig:microstructure}
\end{figure}

The microstructure is represented by an ensemble of $100$ periodic volume elements, each of which comprises $N = 100 \times 100$ square cells. Each individual cell is randomly assigned the properties of one of the two phases (hard or soft), such that the volume fraction of hard phase matches a target value $\varphi_\mathrm{hard}$ for each volume element; the same thus also holds for the ensemble. Below, a reference case is considered for which $\varphi_\mathrm{hard} = 0.25$. A typical volume element from this ensemble is shown in Figure~\ref{fig:microstructure}, wherein the hard cells are shown in red and the soft cells in blue; the periodicity is indicated with dashed lines.

Consistent with the morphological approximation -- the square cells -- the resolution at which we consider the quantities of interests is that of the individual cells. Local (tensorial) quantities are thus considered as averages of each cell. As discussed in more detail below, this introduces the cell size as the governing length-scale within the microstructure. This length-scale implicitly regularizes the damage evolution, thereby rendering it independent of the underlying numerical discretization.

The response is computed using the finite element method. Four bi-quadratic finite elements are used to discretize each cell (see Figure~\ref{fig:microstructure}). Reduced integration is applied, corresponding to four Gauss-points per finite element. For this discretization, the considered (cell averaged) quantities are converged within a 5\% relative error with respect to a much finer discretization. Using hexagons, it was also found that the mechanics do not show a pronounced dependence on the square shape of the cells \citep{DeGeus2015}.

\subsection{Material model}

Both phases are assumed isotropic elasto-plastic, whereby the two phases only differ through a different yield stress and hardening modulus. The model due to \citet{Simo1992a} is used, which properly accounts for the large deformations that are present at fracture. The Kirchhoff stress $\bm{\tau}$ is thereby linearly related to the logarithmic elastic strain $\bm{\varepsilon}_\mathrm{e}$, using the neo-Hookean elasticity tensor parametrized using Young's modulus $E$ and Poisson's ratio $\nu$. $J_2$-plasticity is used, whereby linear hardening is assumed. This corresponds to the following yield function
\begin{equation}
  \Phi( \bm{\tau}, \varepsilon_\mathrm{p} ) =
  \tau_\mathrm{eq} (\bm{\tau}) - (\tau_\mathrm{y0} + H \varepsilon_\mathrm{p})
  \leq 0
  \qquad \mathrm{with} \quad
  \varepsilon_\mathrm{p} = \int\limits_0^t \dot{\varepsilon}_\mathrm{p} \;\mathrm{d} t'
\end{equation}
wherein $\tau_\mathrm{eq}$ is the equivalent Kirchhoff stress and $\varepsilon_\mathrm{p}$ is the accumulated plastic strain. The material model is numerically integrated in pseudo-time using an implicit scheme.

The contrast between the phases is introduced in the yield stress and the hardening modulus of the hard phase, which are a factor $\chi$ higher compared to the soft phase, i.e.
\begin{equation}
  \tau_\mathrm{y0}^\mathrm{hard}
  = \chi \; \tau_\mathrm{y0}^\mathrm{soft}
  \qquad
  H^\mathrm{hard}
  = \chi \; H^\mathrm{soft}
\end{equation}
wherein $\chi = 2$ is used as reference value. The other parameters are chosen in a regime that is representative for dual-phase steel \cite{Sun2009, Al-Abbasi2003, Asgari2009, Vajragupta2012}:
\begin{equation}
  \frac{\tau_\mathrm{y0}^\mathrm{soft}}{E} = 3 \cdot 10^{-3} \qquad
  \frac{H^\mathrm{soft}}{E} = 4 \cdot 10^{-3} \qquad
  \nu = 0.3
\end{equation}
It is emphasized that the elasto-plastic response is computed at the sub-cell level, i.e.\ in the integration points of the finite element discretization in Figure~\ref{fig:microstructure}. However, stresses and plastic strains are subsequently averaged per cell for visualization and to compute the damage -- see below.

\subsection{Damage and fracture}

The soft phase is assumed to fail by ductile fracture. In the first part of the paper the hard phase is assumed not to fail. Earlier work \citep{DeGeus2015b} has shown this to be a reasonable assumption for the pure shear deformation considered here. In Section~\ref{sec:triax}, where we study the influence of stress triaxiality, failure of the hard phase is included. Note that, besides Ref.~\citep{DeGeus2015b} on initiation, using the reference case in Section~\ref{sec:reference} it has been verified that failure only occurs in the soft phase also when damage propagation is considered.

In the spirit of the idealized microstructure, damage and fracture are modeled in a crude but efficient way using cell erosion. A damage indicator, $D$, is used to signal the fracture of a single cell when $D = 1$. The ductile Johnson-Cook damage indicator \citep{Johnson1985} is used for the soft phase. Reformulated in an incremental form, it compares the incrementally accumulated plastic strain, $\dot{\varepsilon}_\mathrm{p}$, to a critical value, $\varepsilon_\mathrm{c}$, which depends on the stress state in the cell. In particular,
\begin{equation}\label{eq:model:D}
  D =
  \int_0^t \frac{\dot{\varepsilon}_\mathrm{p}}{\varepsilon_\mathrm{c} (\eta)}
  \; \mathrm{d} t^\prime
  \qquad
  \text{with}
  \qquad
  \varepsilon_\mathrm{c} = A \exp \left( - B \eta \right) + \varepsilon_\mathrm{pc}
\end{equation}
in which $\eta$ is the (average) stress triaxiality in the cell. The parameters $A$, $B$, and $\varepsilon_\mathrm{pc}$ are taken loosely representative for dual-phase steel \cite{Vajragupta2012,DeGeus2015b}:
\begin{equation}
  A^\mathrm{soft} = 0.2 \quad
  B^\mathrm{soft} = 1.7 \quad
  \varepsilon_\mathrm{pc}^\mathrm{soft} = 0.1
\end{equation}
Note that the damage is `non-local' as $\dot{\varepsilon}_\mathrm{p}$ and $\eta$ are cell averaged quantities and $D$ is computed on a cell-by-cell basis only. The cell size thereby enters as the governing length-scale.

As long as $D < 1$ the cell continuously hardens. Then, when $D = 1$, the cell is instantaneously removed from the microstructure without any prior softening. This is done for all cells in the microstructure in which $D = 1$ at that instant. I.e.\ multiple voids can appear simultaneously. This thus corresponds to a semi-implicit temporal discretization, in which the plasticity and damage are integrated implicitly but the cell erosion is not reversible within an increment. Because of the latter, the deformation is applied in small increments (see below). As a result, voids mostly nucleate one by one.

More in detail, within each increment, a two-stage algorithm is used to iteratively establish a state that satisfies both mechanical equilibrium and the fracture criterion. First, within each increment, regular equilibrium iterations are performed for the microstructural state established in the previous increment. Subsequently, any cell for which $D \geq 1$ is removed from the microstructure. Furthermore, single node connections between two elements, which can arise as a consequence of the square cells, are disconnected. If one or more cells are removed, mechanical equilibrium is re-established by additional equilibrium iterations, whereby the stress on the newly created free surface is relaxed in small steps to avoid convergence issues. Convergence of the numerical solution scheme is guaranteed as long as these unloading steps are sufficiently small. Here, the amplitude of the force has been exponentially reduced to zero in $10$ \textit{equilibrium increments}. In practice this makes the simulation between two and four times slower than for the case than without any damage, as in most increments no voids are nucleated. Note that the process of cell erosion is repeated until $D < 1$ for all remaining cells in the microstructure, before proceeding to the next deformation increment.

\subsection{Applied deformation}

Since the microscopic failure is driven by plasticity, macroscopic shear deformation is applied. Pure shear is applied in conjunction with a plane strain condition. This corresponds to the following macroscopic logarithmic strain tensor:
\begin{equation}\label{eq:model:deformation}
  \bar{\bm{\varepsilon}} =
  \tfrac{\sqrt{3}}{2} \, \bar{\varepsilon}\,
  \left(
    \vec{e}_\mathrm{x} \vec{e}_\mathrm{x} -
    \vec{e}_\mathrm{y} \vec{e}_\mathrm{y}
  \right)
\end{equation}
where $\bar{\varepsilon}$ is the macroscopic equivalent strain. This applied deformation is isochoric. Any local volumetric deformation, and resulting hydrostatic stress, is therefore a consequence of the microstructural heterogeneity. The deformation is applied in increments of $\dot{\bar{\varepsilon}} = 0.0001$ up to the moment of global instability. Beyond this point, the response is affected by the assumed periodicity and therefore loses its physical relevance.

We define the fraction strain $\bar{\varepsilon}_\mathrm{f}$ as the applied strain at which the stress has dropped $1\%$ below its peak value, i.e.\
\begin{equation}\label{eq:model:maxstress}
  \bar{\tau}_\mathrm{eq} ( \bar{\varepsilon}_\mathrm{f} ) =
  0.99 \, \max \big[
   \bar{\tau}_\mathrm{eq} (\bar{\varepsilon} \leq \bar{\varepsilon}_\mathrm{f})
  \big]
\end{equation}
wherein $\bar{\tau}_\mathrm{eq}$ is the macroscopic (periodic volume averaged) equivalent Kirchhoff stress.

\section{Reference case}
\label{sec:reference}

\subsection{Macroscopic response}
\label{sec:reference:macro}

\begin{figure}[htbp]
  \centering
  \includegraphics[width=0.5\textwidth]{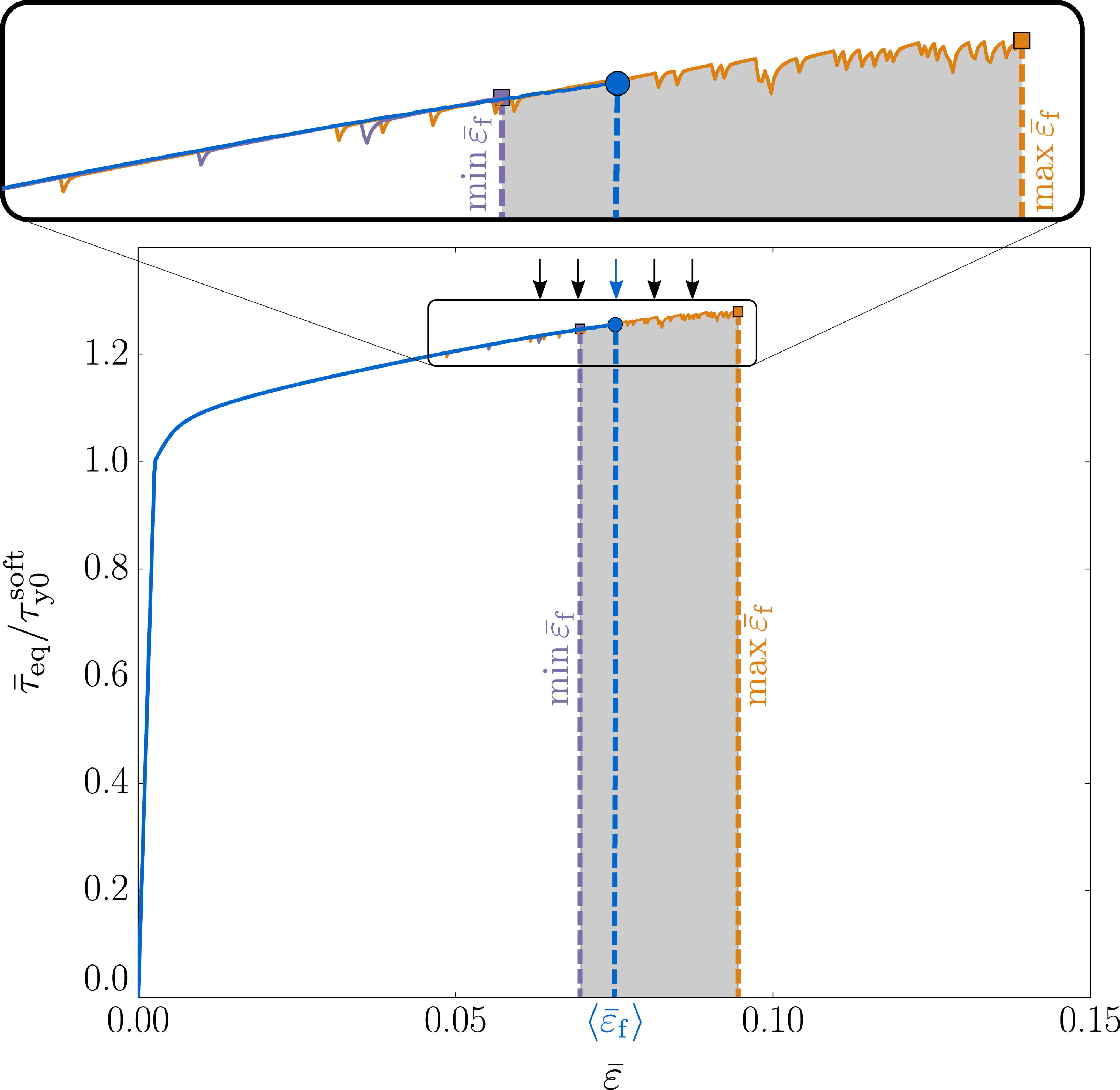}
  \caption{Macroscopic response of the reference ensemble ($\varphi_\mathrm{hard} = 0.25$ and $\chi = 2$) for the microstructures with the lowest (purple line) and the highest fracture strain (orange line) and the homogenized response (blue line), with the resulting fracture strain $\langle \bar{\varepsilon}_\mathrm{f} \rangle$. $\bar{\tau}_\mathrm{eq}$ is the equivalent stress and $\bar{\varepsilon}$ is the (applied) equivalent strain. The arrows indicate the different snapshots shown in Figure~\ref{fig:config013:hotspot_delta-D}.}
  \label{fig:config013:macroscopic}
\end{figure}

For the reference ensemble, with hard phase volume fraction $\varphi_\mathrm{hard} = 0.25$ and phase contrast $\chi = 2$, the macroscopic response is shown in Figure~\ref{fig:config013:macroscopic}. The microstructural volume elements that exhibit the lowest and the highest fracture strain are included using purple and orange lines respectively, each of which is truncated at its fracture strain $\bar{\varepsilon}_\mathrm{f}$. The range of fracture strains of the different microstructures is highlighted in gray. A \textit{homogenized response} is introduced for comparison purposes as the ensemble average of all $100$ realizations (blue line). To be independent of a single realization but still sensitive to the extreme cases, the homogenized fracture strain $\langle \bar{\varepsilon}_\mathrm{f} \rangle$ is not defined as the average or the lowest fracture strain, but as the strain at which $5\%$ of the volume elements in the ensemble have failed (marked with the blue dot).

As expected, Figure~\ref{fig:config013:macroscopic} reveals that the elastic and the initial plastic responses of all volume elements virtually coincide, since the hard phase volume fraction of all volume elements is identical. At $\bar{\varepsilon} = 0.044$ the first void of the ensemble nucleates, as is observed in the zoom in Figure~\ref{fig:config013:macroscopic}. Due to the adopted cell erosion, the microstructure unloads elastically, which is macroscopically observed as a sharp decrease in stress. Upon further deformation, the cells in the microstructure start to yield again and the hardening prior to void nucleation is resumed, until the next void nucleates. The most striking observation to be made from Figure~\ref{fig:config013:macroscopic} is the large range of strains in which the individual microstructures fail. This suggests a strong influence of the phase distribution, since this is the only feature discriminating them. Furthermore, one observes that the voids have little effect on the macroscopic response up to the point of instability.

\begin{figure}[htp]
  \centering
  \includegraphics[width=0.523\textwidth]{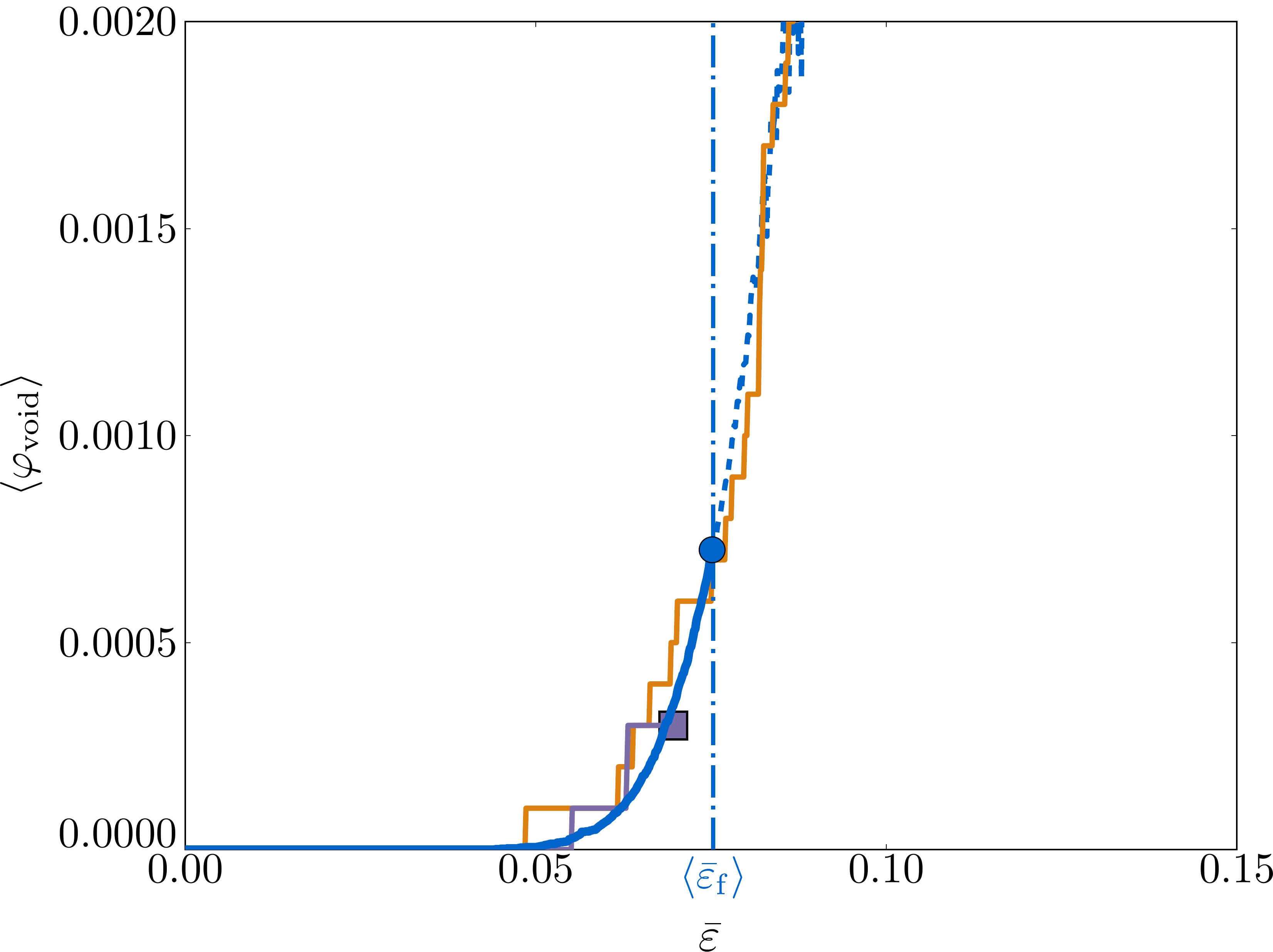}
  \caption{Ensemble averaged void volume fraction, $\langle \varphi_\mathrm{void} \rangle$, as a function of the applied equivalent strain, $\bar{\varepsilon}$. The homogenized fracture strain, $\langle \bar{\varepsilon}_\mathrm{f} \rangle$, is indicated in blue. The void volume fractions of the microstructures with the lowest and the highest fracture strain are also indicated (purple and orange line respectively).}
  \label{fig:config013:voids}
\end{figure}

The ensemble averaged void volume fraction, $\langle \varphi_\mathrm{void} \rangle$, is shown in blue in Figure~\ref{fig:config013:voids} as a function of the applied strain, $\bar{\varepsilon}$. It increases exponentially with $\bar{\varepsilon}$. The amount of voids is significantly higher in those microstructures that fracture at a high strain, than in those that fracture at a low strain -- as is illustrated by the purple and orange lines, corresponding to individual microstructures with the lowest and the highest fracture strain respectively. In other words, in some microstructures many voids may nucleate throughout the deformation which do not cause catastrophic fracture, whereas in other microstructures just a few voids are enough to trigger final fracture. This substantiates the presumption of the strong influence of the phase distribution.

\subsection{Typical microscopic responses}

\begin{figure}[htp]
  \centering
  \includegraphics[width=1.\textwidth]{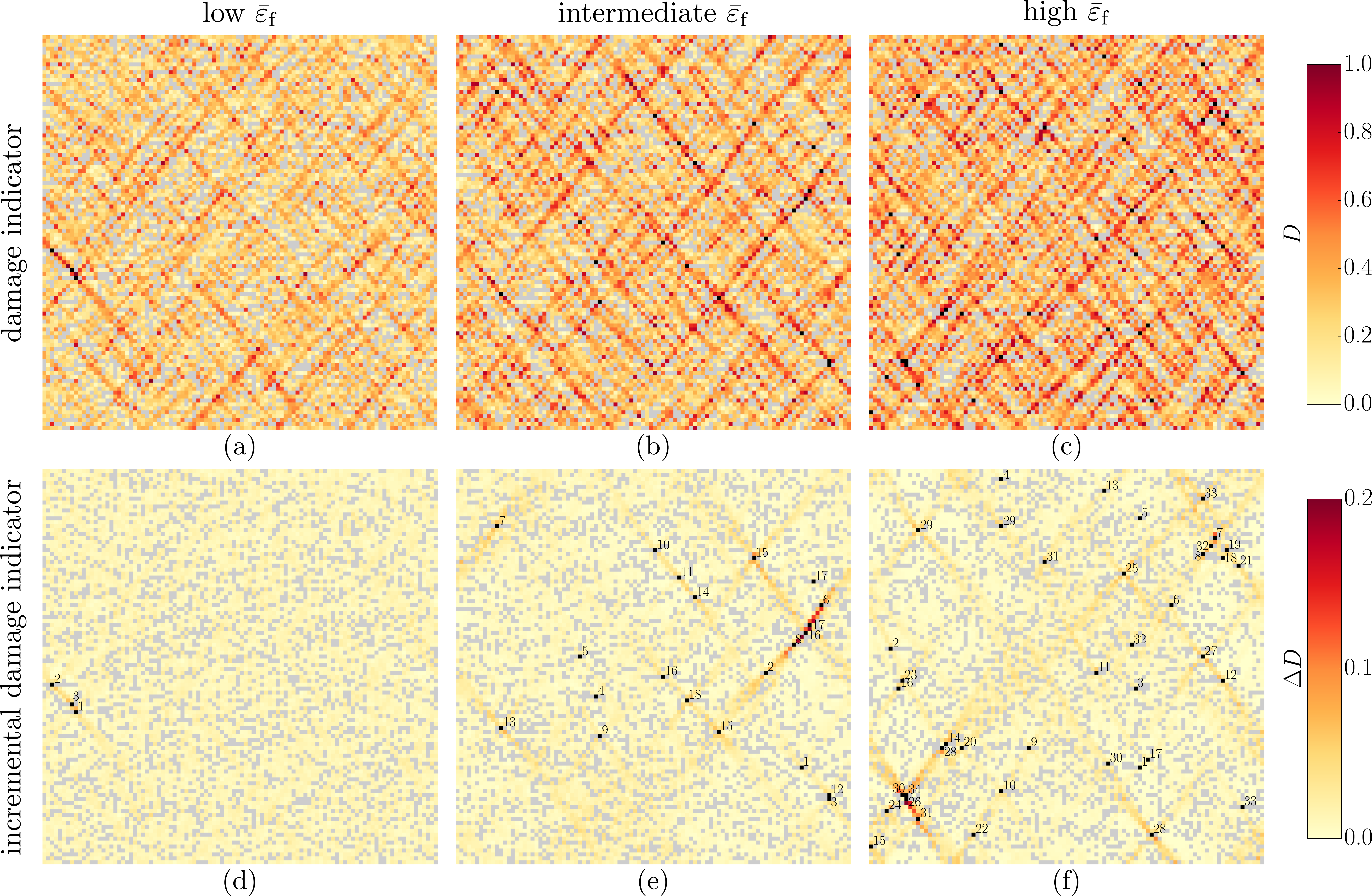}
  \caption{Response of three individual microstructures taken from the reference ensemble ($\varphi_\mathrm{hard} = 0.25$ and $\chi = 2$), at their respective fracture strains, $\bar{\varepsilon}_\mathrm{f}$. Top: the damage indicator $D$, bottom: the incremental damage indicator $\Delta D$. The hard cells are gray, the voids are black. The numbers indicate the sequence of nucleation, 1 nucleates first, then 2, etc. The coordinate system is equal to Figure~\ref{fig:microstructure}.}
  \label{fig:config013:typical}
\end{figure}

To further illustrate the response of the reference ensemble, the microscopic damage responses of three individual microstructures are plotted in Figure~\ref{fig:config013:typical} -- the ones with the lowest, the median, and the highest fracture strain of the ensemble. The top row shows the distribution of damage $D$ at the point of instability -- shown on the undeformed geometry -- and the bottom row shows the incremental damage $\Delta D$ at that point, which is used to indicate the largest damage activity $D(\bar{\varepsilon}) - D(\bar{\varepsilon}-0.002)$. Since damage is only considered in the soft phase, the hard phase is gray in Figure~\ref{fig:config013:typical}; the voids are black.

In all cases, the damage, $D$, is high in bands at $\pm 45$ degrees angles with respect to the tensile axis (in horizontal direction). These directions correspond to the directions of maximum shear. Note that $D$ is on average higher in Figure~\ref{fig:config013:typical}(c) than in Figure~\ref{fig:config013:typical}(a) because of the higher applied strain. In the microstructure of Figure~\ref{fig:config013:typical}(a), a few voids have become critical, i.e.\ triggered instability, at an early state of the deformation; a more widespread distribution is observed in the other two microstructures.

To focus on localization, the incremental damage, $\Delta D$, is inspected (bottom row of Figure~\ref{fig:config013:typical}). A more pronounced pattern is observed in this case, whereby the highest values of $\Delta D$ are located in bands that connect two or more voids (in black) at $\pm 45$ degrees of each other. Within these bands, the highest values of $\Delta D$ occur directly next to the voids. $\Delta D$ also appears to depend on the distance between the voids and the presence of a continuous percolation path connecting them through the soft phase. This is supported by the sequence of void nucleation, indicated with numbers in Figures~\ref{fig:config013:typical}(d--f). The voids in the localization bands are often nucleated in between two existing voids, for example voids \{1--3\} in Figure~\ref{fig:config013:typical}(d), voids \{6,8,16,17\} in Figure~\ref{fig:config013:typical}(e) and voids \{26,30,34\} in Figure~\ref{fig:config013:typical}(f). What appears to make the difference between the microstructure which fails early and the other microstructures is that the first nucleated voids happen to be positioned `favorably' with respect to each other and thus start to interact immediately.

\subsection{Microscopic spatial correlation}
\label{sec:spatial_correlation}

The above observations may be extended to the entire ensemble of $100$ volume elements, and rendered more quantitative by determining spatial correlations. In earlier work, we have found that the average arrangement of the phases around high damage provides an objective measure to identify critical features that are typically found around fracture initiation \citep{DeGeus2015a}. In the context of propagation, this analysis needs further refinement. One particular question concerns the sequence of nucleation events. Or, more precisely, what is the probability of finding hard phase, soft phase, and/or voids at a certain position relative to cells with a high increment of damage. The latter thereby acts as an indicator for the location of the next void, or at least for evolving damage. At low $\bar{\varepsilon}$, it provides a measure for the likelihood of void initiation. At the onset of localization, it is expected to be characteristic for the mechanism driving propagation of the damage.

The correlation measures used in this study are explained below for a single volume element. The ensemble averaged equivalents trivially follow. Three different indicator functions are introduced, respectively for the hard phase, the soft phase, and the voids:
\begin{equation}\label{eq:cor:I}
  \mathcal{I}_\mathrm{hard} (\vec{x},\bar{\varepsilon}) =
  \begin{cases}
    1 \;\; \text{if}\; \vec{x} \in \text{hard} \\
    0 \;\; \text{otherwise}
  \end{cases}
  \quad
  \mathcal{I}_\mathrm{soft} (\vec{x},\bar{\varepsilon}) =
  \begin{cases}
    1 \;\; \text{if}\; \vec{x} \in \text{soft} \\
    0 \;\; \text{otherwise}
  \end{cases}
  \quad
  \mathcal{I}_\mathrm{void} (\vec{x},\bar{\varepsilon}) =
  \begin{cases}
    1 \;\; \text{if}\; \vec{x} \in \text{void} \\
    0 \;\; \text{otherwise}
  \end{cases}
\end{equation}
which together uniquely describe the current microstructure for a given applied macroscopic strain $\bar{\varepsilon}$. To be precise, for a point $\vec{x}$ only one of these three indicators equals one, while the other two are equal to zero. Their spatial averages are respectively the (current) hard phase volume fraction $\varphi_\mathrm{hard}$, the soft phase volume fraction $\varphi_\mathrm{soft}$, and the void volume fraction $\varphi_\mathrm{void}$. The probability of recovering hard phase at a certain distance $\Delta \vec{x}$ relative to (high) incremental damage can now be computed as the following weighted spatial average
\begin{equation}\label{eq:cor:P}
  \mathcal{P}_{(\Delta D \star \mathcal{I}_\mathrm{hard})} ( \Delta \vec{x} )
  =
  \frac{
    \sum_{i=1}^N\;
    \Delta D (\vec{x}_i) \;
    \mathcal{I}_\mathrm{hard} ( \vec{x}_i + \Delta \vec{x} )
  }{
    \sum_{i=1}^N\;
    \Delta D (\vec{x}_i) \hfill
  }
\end{equation}
wherein $i$ loops over all $N$ cells in the volume element. Its value is interpreted as follows. If there is no correlation between a high value for the incremental damage and the hard phase at a certain position $\Delta \vec{x}$ relative to it, then $\mathcal{P}_{(\Delta D \star \mathcal{I}_\mathrm{hard})} (\Delta \vec{x}) = \varphi_\mathrm{hard}$, i.e.\ it is equal to the probability of finding the hard phase anywhere. A value $\mathcal{P}_{(\Delta D \star \mathcal{I}_\mathrm{hard})} (\Delta \vec{x}) > \varphi_\mathrm{hard}$ corresponds to an elevated probability of finding hard phase at the position $\Delta \vec{x}$ relative to a cell with high incremental damage, and $\mathcal{P}_{(\Delta D \star \mathcal{I}_\mathrm{hard})} (\Delta \vec{x}) < \varphi_\mathrm{hard}$ implies that it is less likely to find the hard phase there.

The same computation and interpretation holds for the soft phase and the voids, once the indicator function $\mathcal{I}$ and volume fraction $\varphi$ are substituted correspondingly. The ensemble averages are finally determined by averaging over all volume elements in the ensemble.

It is observed from Figure~\ref{fig:config013:voids} that the volume fraction of voids $\langle \varphi_\mathrm{void} \rangle \ll 1$. This immediately implies that $\langle \mathcal{P}_{(\Delta D \star \mathcal{I}_\mathrm{soft})} \rangle \approx 1 - \langle \mathcal{P}_{(\Delta D \star \mathcal{I}_\mathrm{hard})} \rangle$, and thus that $\mathcal{P}_{(\Delta D \star \mathcal{I}_\mathrm{hard})} (\Delta \vec{x}) < \varphi_\mathrm{hard}$ corresponds to an elevated probability of having the soft phase there. $\langle \mathcal{P}_{(\Delta D \star \mathcal{I}_\mathrm{soft})} \rangle$ is therefore not shown below.

Figure~\ref{fig:config013:hotspot_delta-D} shows the probability of finding hard phase as a function of the position relative to high incremental damage (top row), and that of finding voids (bottom row) at that relative position. The horizontal and vertical directions thus coincide with Figures~\ref{fig:microstructure} and~\ref{fig:config013:typical}. The chosen color-scales maximize the interpretability of the results, whereby their ranges are in the order of the hard phase and void volume fraction respectively. In the top row, red corresponds to an elevated probability of having the hard phase while blue corresponds to a lower than average probability of hard phase and hence an elevated probability of finding the soft phase. In the bottom row, red corresponds to an elevated probability of having voids at that position relative to evolving damage, while white implies that it is unlikely to find voids there. As the incremental damage is only defined outside the voids the probability of the central cells on the bottom row is always zero. The columns, finally, correspond to different levels of applied strain -- which increases from left to right -- as indicated by the arrows in Figure~\ref{fig:config013:macroscopic}. The middle column corresponds to the homogenized fracture strain, $\langle \bar{\varepsilon}_\mathrm{f} \rangle$.

The results indicate a consistent preferential arrangement of the hard and soft phase around high values of the incremental damage throughout the deformation history (cf.\ Figures~\ref{fig:config013:hotspot_delta-D}(a--e)). Voids are likely to nucleate there where regions of the hard phase, aligned along the tensile axis, are intersected by regions of the soft phase at approximately $\pm 45$ degree angles, i.e.\ aligned with the directions of maximum shear. Simple mechanical arguments explain these features. The presence of the hard phase to the left and right triggers a state of high hydrostatic tensile stress, while the soft bands facilitate plasticity \citep{DeGeus2015a}. The combined effect makes such a configuration a `damage hot-spot'.

The void volume fraction increases with increasing strain, as observed from the background color in Figures~\ref{fig:config013:hotspot_delta-D}(f--j). A clear correlation is observed with existing voids in regions also at approximately $\pm 45$ degree angles, whereby the probability is highest close to the site with high incremental damage in the center. It is unlikely to find voids in the tensile direction. Combined with the observations on the arrangement of the phases, this suggests that the damage localizes in bands of the soft phase at approximately $\pm 45$ degree angles. Several voids that have nucleated in initiation `hot-spots' are thereby interlinked. Voids thus nucleate because of a local critical phase distribution. The propagation of damage is controlled on a longer length-scale. The critical configuration is found to be several initiation `hot-spots' that are close to each other in the directions of shear, with a percolation path through the soft phase connecting them. This explains why the microstructure in Figure~\ref{fig:config013:typical}(d) fails at a lower strain than the one in Figure~\ref{fig:config013:typical}(f).

A strong dependence on the microstructure for the sequential nucleation of voids is thus confirmed (Figures~\ref{fig:config013:hotspot_delta-D}(a--e)). The fact that some microstructures fail much earlier than others depends on the relative location of subsequently nucleated voids (cf.\ Figures~\ref{fig:config013:hotspot_delta-D}(f--j) to Figures~\ref{fig:config013:typical}(d--f)). If the voids nucleate sufficiently far away and/or are not aligned with the shear, they do not propagate to final localization.

\begin{figure}[htp]
  \centering
  \includegraphics[width=1.0\textwidth]{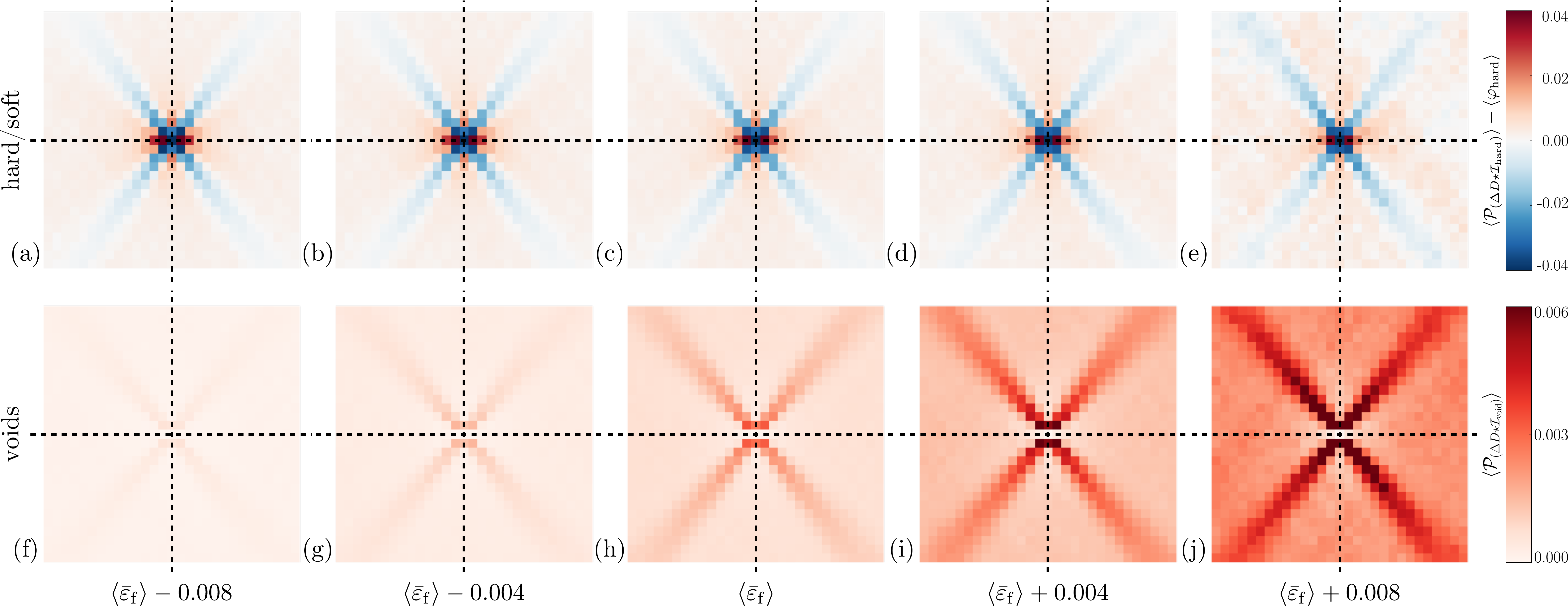}
  \caption{Probability of the hard phase (and the soft phase) ($\mathcal{P}_{(\Delta D \star \mathcal{I}_\mathrm{hard})}$, top) and of the voids ($\mathcal{P}_{(\Delta D \star \mathcal{I}_\mathrm{void})}$, bottom) at a distance $\Delta \vec{x}$ relative to high incremental damage, in the center. The coordinate system is equal to Figure~\ref{fig:microstructure}. All results are for the reference ensemble ($\varphi_\mathrm{hard} = 0.25$ and $\chi = 2$).}
  \label{fig:config013:hotspot_delta-D}
\end{figure}

\section{Parameter variation}
\label{sec:parameter}

\subsection{Effect of hard phase volume fraction}

\begin{figure}[htp]
  \centering
  \includegraphics[width=0.5\textwidth]{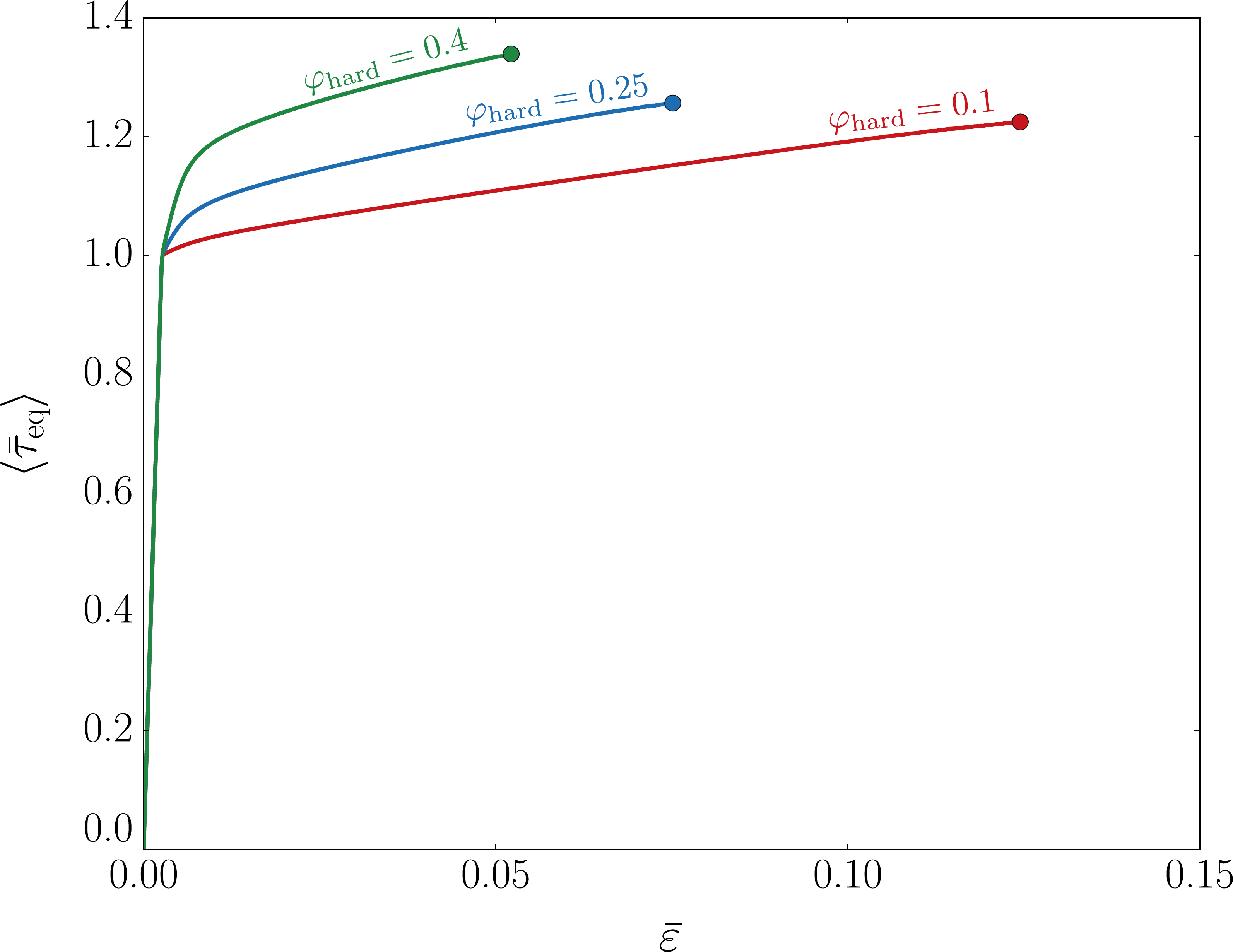}
  \caption{Homogenized macroscopic response of three different ensembles, respectively with hard phase volume fraction $\varphi_\mathrm{hard} = 0.1$, $0.25$, and $0.4$; $\langle \bar{\tau}_\mathrm{eq} \rangle$ is the homogenized macroscopic equivalent stress and $\bar{\varepsilon}$ is the homogenized equivalent strain.}
  \label{fig:varphi:macro}
\end{figure}

Three different hard phase volume fractions are compared for the reference phase contrast $\chi = 2$. In addition to the reference ensemble, $100$ volume elements with $\varphi_\mathrm{hard} = 0.1$ and $0.4$ are used. The homogenized macroscopic responses for the three ensembles are shown in Figure~\ref{fig:varphi:macro} using different colors for the different hard phase volume fractions. Increasing the volume fraction leads to a macroscopically higher initial hardening modulus, while the hardening rate at larger strains is approximately the same for all cases. At the same time the fracture strain $\langle \bar{\varepsilon}_\mathrm{f} \rangle$ decreases with increasing $\varphi_\mathrm{hard}$, while the fracture stress $\langle \bar{\tau}_\mathrm{f} \rangle$ increases to some extent with $\varphi_\mathrm{hard}$. Indeed, the soft phase needs to accommodate more deformation when the amount of hard phase is increased, and therefore fails sooner at the global level. This reflects the classical strength versus ductility trade-off.

\begin{figure}[htp]
  \centering
  \includegraphics[width=0.523\textwidth]{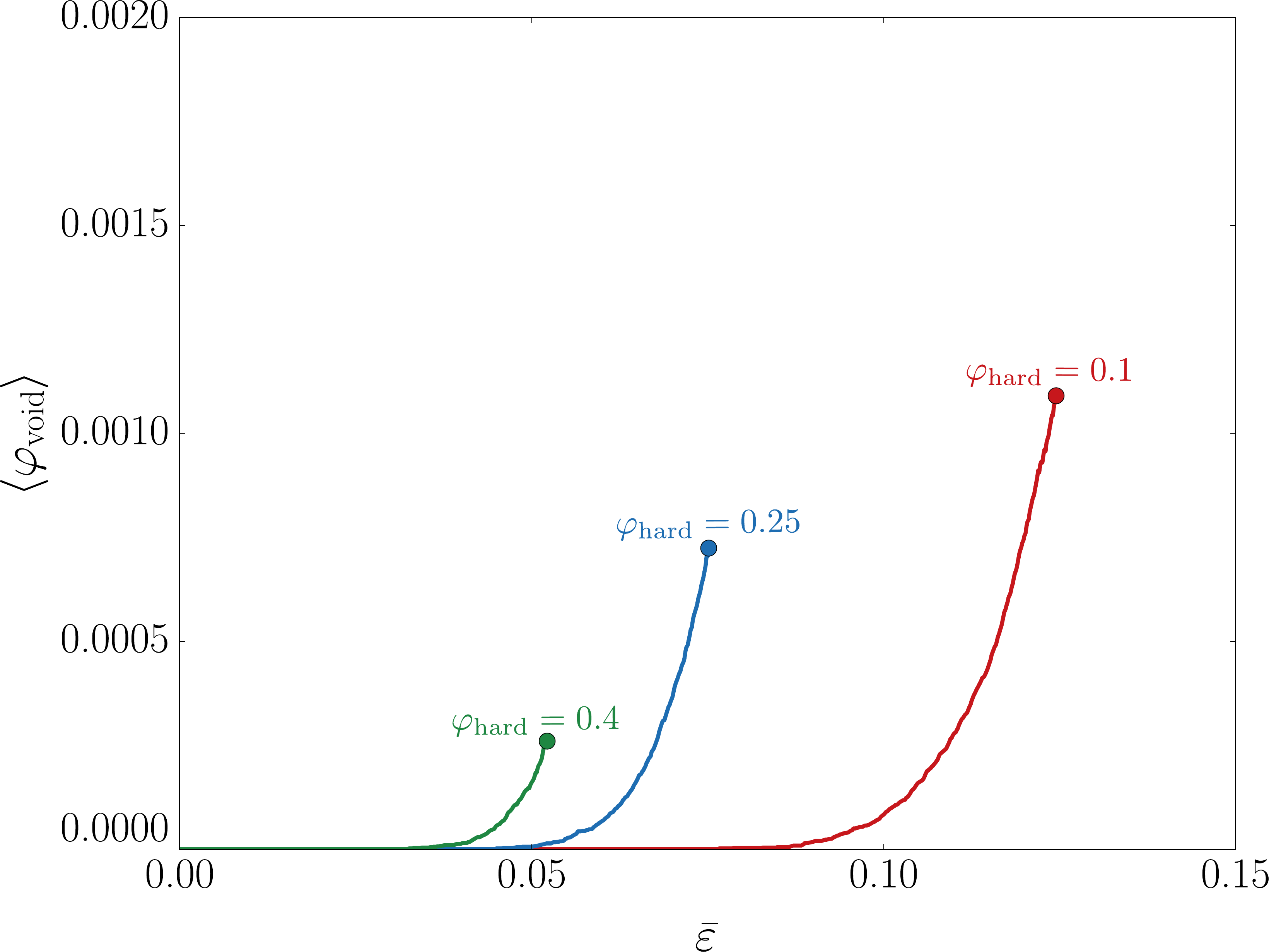}
  \caption{Ensemble averaged void volume fraction, $\langle \varphi_\mathrm{void} \rangle$, as a function of the (applied) equivalent strain, $\bar{\varepsilon}$, for hard phase volume fractions $\varphi_\mathrm{hard} = 0.1$, $0.25$, and $0.4$.}
  \label{fig:varphi:void}
\end{figure}

The evolution of the ensemble average void volume fraction $\langle \varphi_\mathrm{void} \rangle$ is shown in Figure~\ref{fig:varphi:void} for the different hard phase volume fractions. They are all truncated at the homogenized fracture strain $\langle \bar{\varepsilon}_\mathrm{f} \rangle$, as defined in Section~\ref{sec:reference:macro}. In accordance with the macroscopically observed fracture properties, the voids start nucleating at a lower strain for a higher amount of hard phase. The rate at which $\langle \varphi_\mathrm{void} \rangle$ increases is comparable, irrespective of the hard phase volume fraction. Interestingly, however, the void volume fraction at which fracture occurs reduces with an increasing hard phase volume fraction, i.e.\ fewer voids are needed for instability to occur in a microstructure with a higher volume fraction of hard phase. The difference between the lowest and the highest volume fraction is quite substantial: approximately a factor of four.

\begin{figure}[htp]
  \centering
  \includegraphics[width=1.0\textwidth]{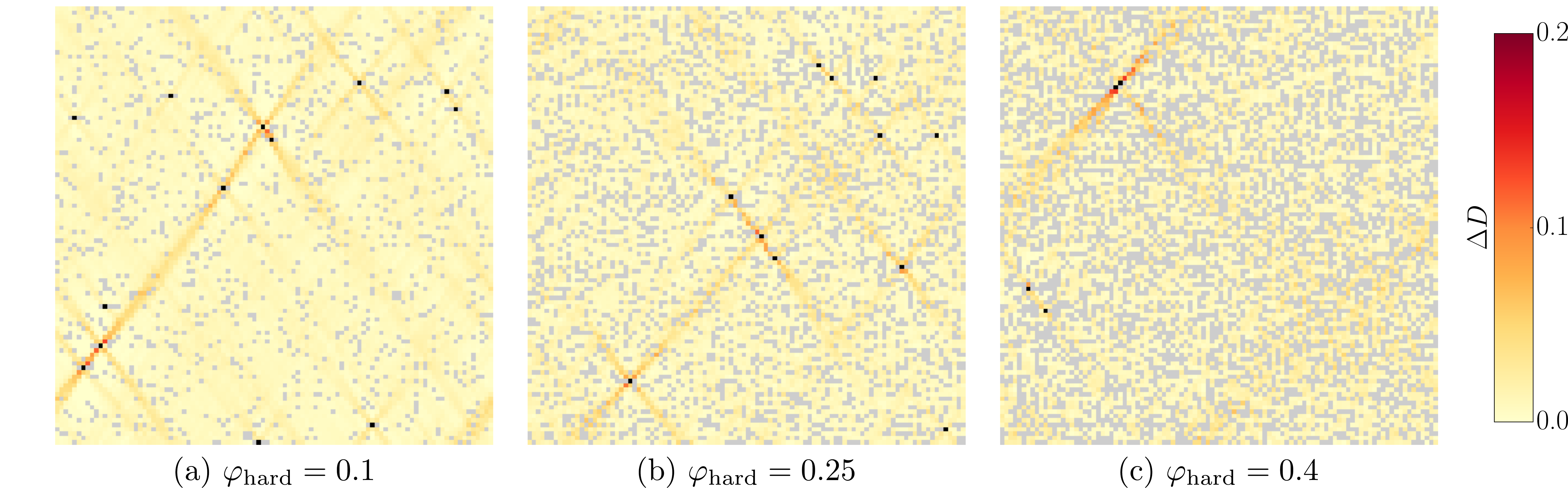}
  \caption{Incremental damage indicator $\Delta D$ of a typical response for hard phase volume fractions $\varphi_\mathrm{hard} = 0.1$, $0.25$, and $0.4$ (for a phase contrast $\chi = 2$), at their respective fracture strains $\bar{\varepsilon}_\mathrm{f} = \langle \bar{\varepsilon}_\mathrm{f} \rangle$. The hard cells are gray, the voids are black. The coordinate system is equal to Figure~\ref{fig:microstructure}.}
  \label{fig:varphi:typical}
\end{figure}

The incremental damage response of three typical microstructures for the different hard phase volume fractions are shown in Figure~\ref{fig:varphi:typical}, where the hard phase volume fraction increases from left to right. For each case, that microstructure is selected whose fracture strain coincides with the homogenized fracture strain, i.e.\ $\bar{\varepsilon}_\mathrm{f} = \langle \bar{\varepsilon}_\mathrm{f} \rangle$. Voids are typically present where the local phase distribution is similar to that suggested by Figures~\ref{fig:config013:hotspot_delta-D}(a--e). High damage activity is observed between voids that are aligned at $\pm 45$ degree angles with respect to the horizontally aligned tensile axis. The value of $\Delta D$ correlates to the number of voids along these directions and their relative distance. For $\varphi_\mathrm{hard} = 0.4$, in Figure~\ref{fig:varphi:typical}(c), the amount of voids is significantly less than for the lower hard phase volume fractions. Localization is triggered where a long percolated band of soft phase is aligned with the shear direction.

\begin{figure}[htp]
  \centering
  \includegraphics[width=0.6\textwidth]{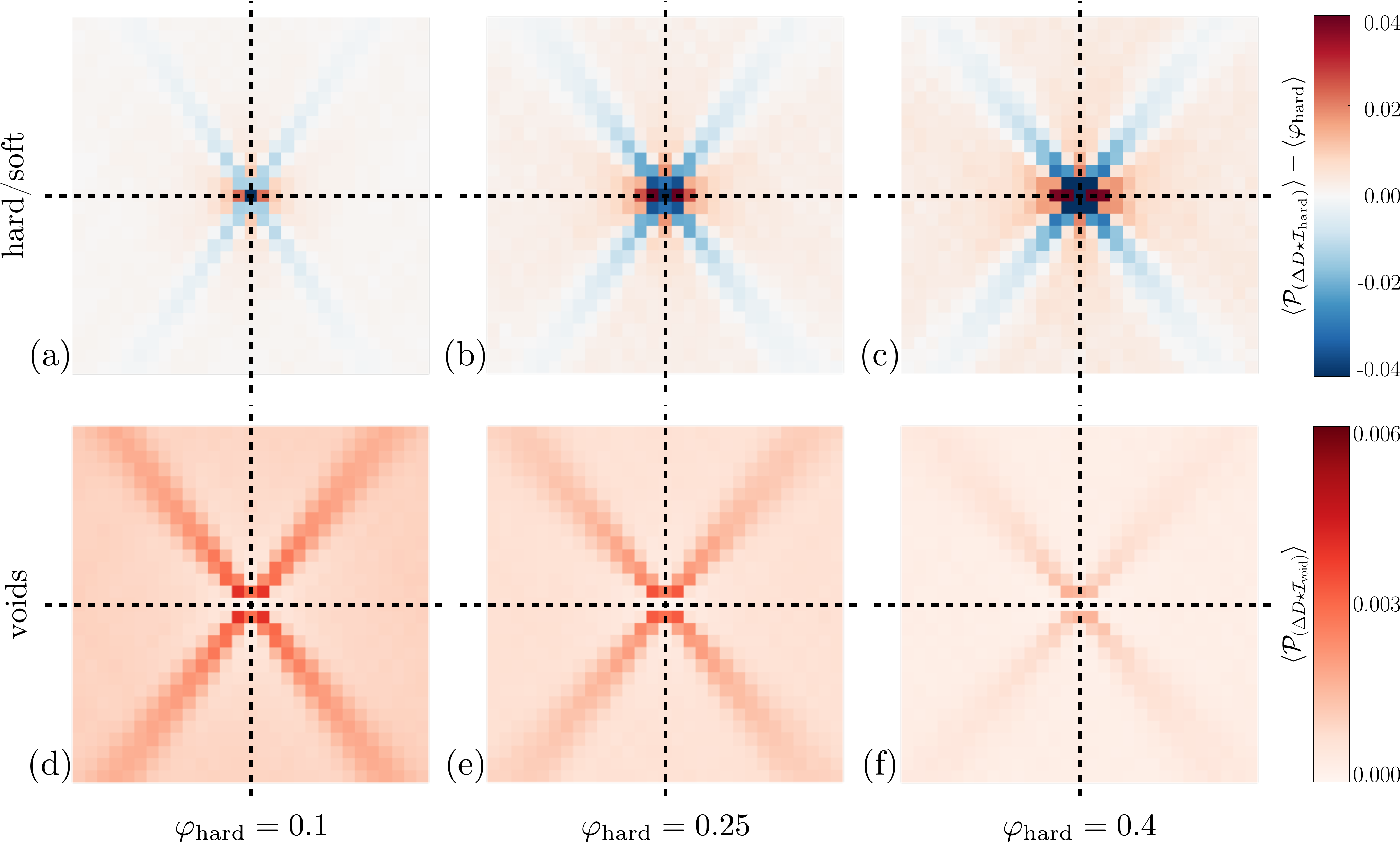}
  \caption{Probability of hard (and soft) ($\mathcal{P}_{(\Delta D \star \mathcal{I}_\mathrm{hard})}$, top) or voids ($\mathcal{P}_{(\Delta D \star \mathcal{I}_\mathrm{void})}$, bottom) at a distance $\Delta \vec{x}$ relative to cells with high incremental damage, in the center. The coordinate system is equal to Figure~\ref{fig:microstructure}. From left to right, the hard phase volume fraction $\varphi_\mathrm{hard} = 0.1$, $0.25$, and $0.4$ respectively.}
  \label{fig:varphi:hotspot_delta-D}
\end{figure}

The correlations that are computed from the ensembles with different volume fractions are depicted in Figure~\ref{fig:varphi:hotspot_delta-D}, each at their respective homogenized fracture strain $\langle \bar{\varepsilon}_\mathrm{f} \rangle$. The evolution with strain is similar to the reference case in Figure~\ref{fig:config013:hotspot_delta-D}, and therefore not shown. For all volume fractions the characteristic features are the same: a critical distribution of phases governs damage initiation (Figures~\ref{fig:varphi:hotspot_delta-D}(a--c)), while damage propagation is governed by voids that are critically aligned with the shear directions. With increasing hard phase volume fraction the influence of the spatial distribution of the phases increases (cf.\ Figures~\ref{fig:varphi:hotspot_delta-D}(a--c)). At the same time, the influence of the voids decreases with increasing hard phase volume fraction (cf.\ Figures~\ref{fig:varphi:hotspot_delta-D}(d--f)). These effects can also be observed from the typical microstructures in Figure~\ref{fig:varphi:typical}. For the relatively low hard phase volume fraction in Figure~\ref{fig:varphi:typical}(a) there are many potential localization bands. The one that becomes critical depends on the presence of voids. On the other hand, when the hard phase volume fraction is high, see Figure~\ref{fig:varphi:typical}(c), localization is triggered in the a single or just a few possible localization bands, which is primarily determined by the microstructure.

\subsection{Effect of phase contrast}

For the reference hard phase volume fraction ($\varphi_\mathrm{hard} = 0.25$) the influence of the phase contrast is investigated next. The following ratios of yield stresses and hardening moduli of the hard versus the soft phase are considered: $\chi = \sqrt{2}$, $2$, and $4$.

\begin{figure}[htp]
  \centering
  \includegraphics[width=0.5\textwidth]{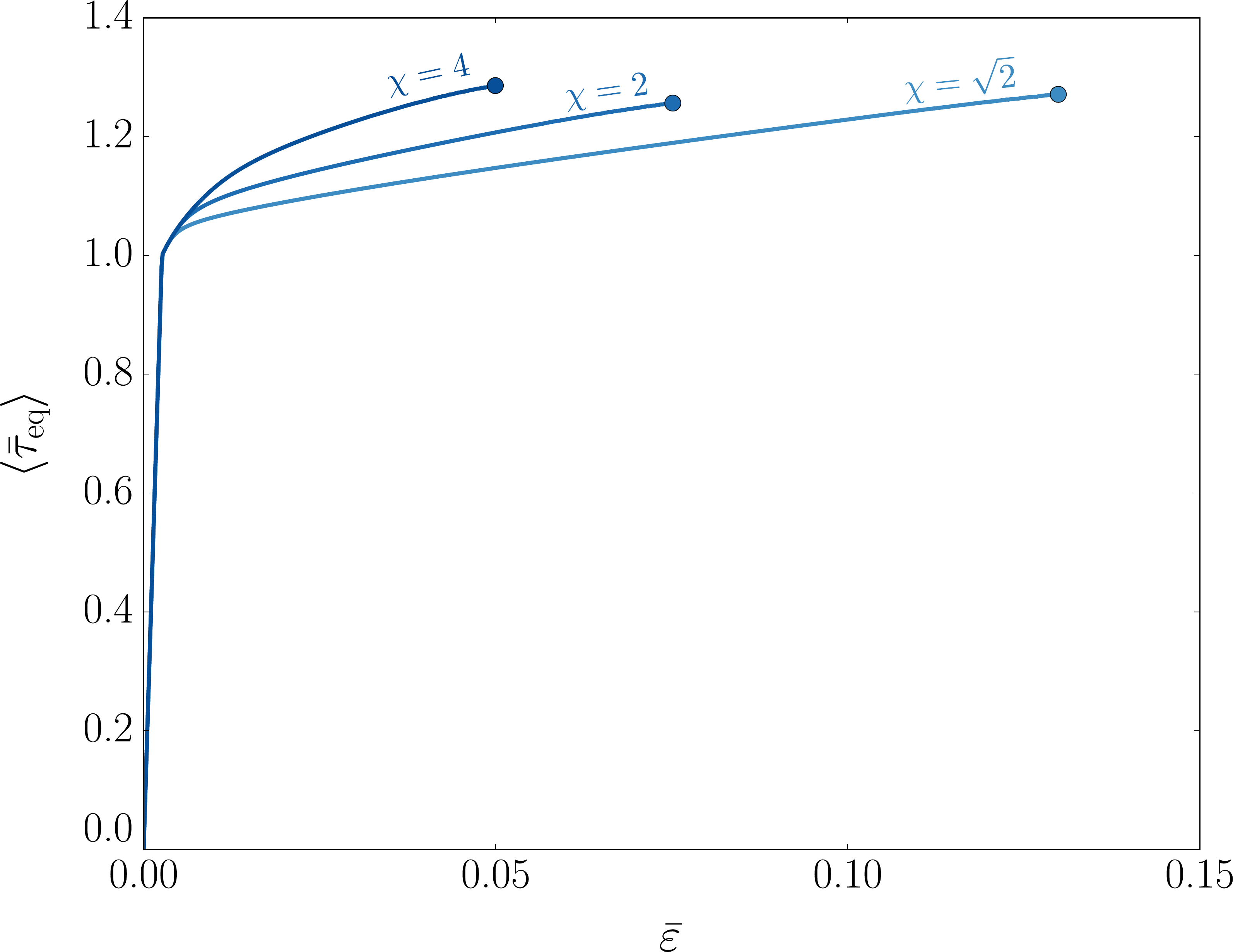}
  \caption{Homogenized macroscopic response of three different ensembles, respectively with phase contrast $\chi = \sqrt{2}$, $2$, and $4$.}
  \label{fig:varchi:macro}
\end{figure}

The homogenized macroscopic responses are shown in Figure~\ref{fig:varchi:macro}, from which it is observed that the macroscopic yield stress is approximately constant, independent of the phase contrast $\chi$, while the hardening increases significantly with the phase contrast. The fracture strain decreases with increasing $\chi$, while the fracture stress increases.

\begin{figure}[htp]
  \centering
  \includegraphics[width=0.523\textwidth]{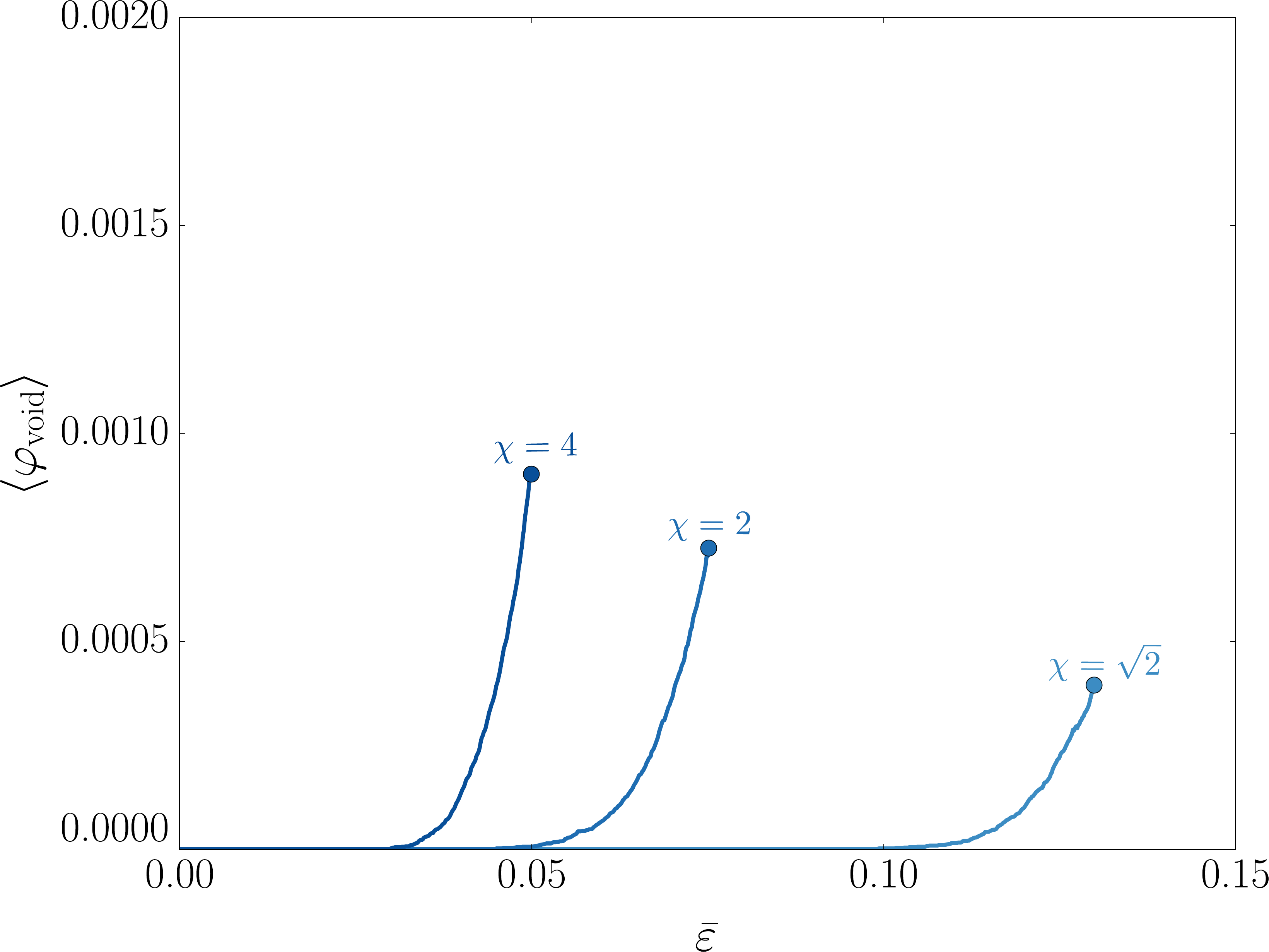}
  \caption{Ensemble averaged void volume fraction, $\langle \varphi_\mathrm{void} \rangle$, as a function of the (applied) equivalent strain, $\bar{\varepsilon}$, respectively for phase contrasts $\chi = \sqrt{2}$, $2$, and $4$.}
  \label{fig:varchi:void}
\end{figure}

The homogenized void volume fraction is shown in Figure~\ref{fig:varchi:void} for the different phase contrasts. Void nucleation starts at a lower strain for higher phase contrast. Moreover, the rate at which $\langle \varphi_\mathrm{void} \rangle$ increases is also higher. The volume fraction at which global failure occurs seems to increase slightly with $\chi$.

\begin{figure}[htp]
  \centering
  \includegraphics[width=1.0\textwidth]{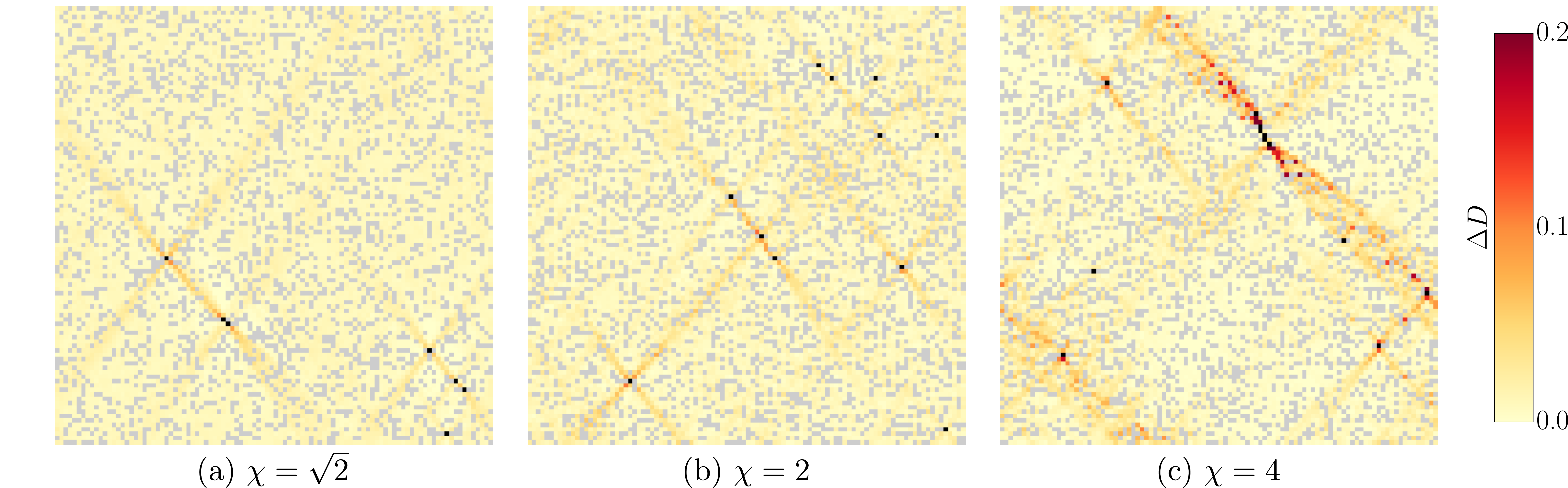}
  \caption{Incremental damage indicator $\Delta D$ of a typical response for phase contrasts $\chi = \sqrt{2}$, $2$, and $4$ (the hard phase volume fraction is that of the reference ensemble, i.e.\ $\varphi_\mathrm{hard} = 0.25$), at their respective fracture strains $\bar{\varepsilon}_\mathrm{f}$. The hard cells are gray, the voids are black. The coordinate system is equal to Figure~\ref{fig:microstructure}.}
  \label{fig:varchi:typical}
\end{figure}

Three typical responses are shown for phase contrasts $\chi = \sqrt{2}$, $2$, and $4$ in Figure~\ref{fig:varchi:typical}. For the lowest phase contrast in Figure~\ref{fig:varchi:typical}(a) the incremental damage is localized in a long percolation path connecting several voids. For $\chi = 4$, in Figure~\ref{fig:varchi:typical}(c) the localization is obviously more influenced by the microstructure. In accordance with Figure~\ref{fig:varchi:void}, more voids are nucleated than for a low phase contrast $\chi$. In this case there is a significant difference between $\Delta D$ around isolated voids and those voids that are in proximity of each other at $\pm 45$ degree angles.

\begin{figure}[htp]
  \centering
  \includegraphics[width=0.6\textwidth]{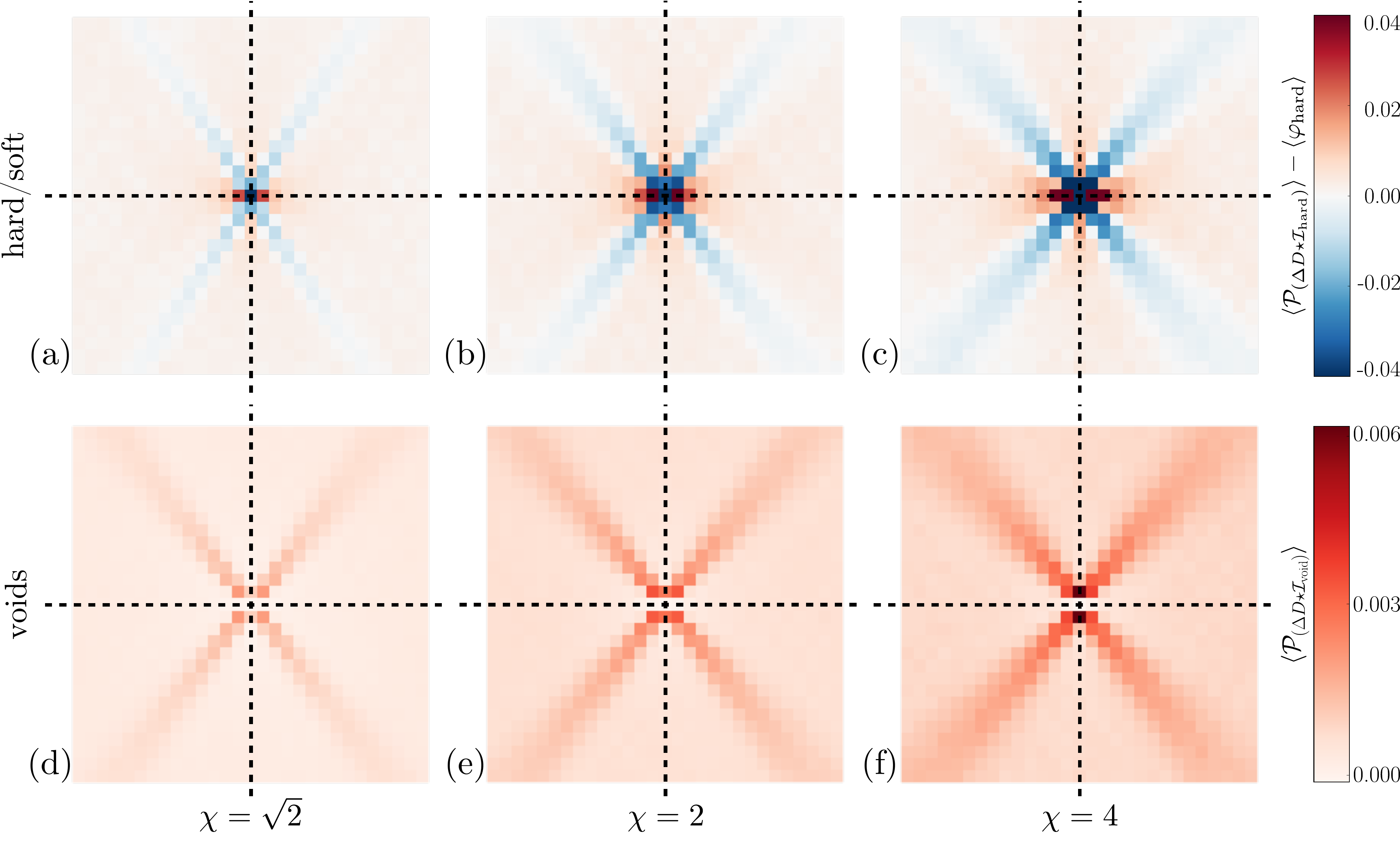}
  \caption{Probability of hard (and soft) ($\mathcal{P}_{(\Delta D \star \mathcal{I}_\mathrm{hard})}$, top) or voids ($\mathcal{P}_{(\Delta D \star \mathcal{I}_\mathrm{void})}$, bottom) at a distance $\Delta \vec{x}$ relative to cells with high incremental damage, in the center. The coordinate system is equal to Figure~\ref{fig:microstructure}. From left to right, the phase contrast $\chi = \sqrt{2}$, $2$, and $4$ respectively.}
  \label{fig:varchi:hotspot_delta-D}
\end{figure}

The average distribution of the hard and the soft phase, and of the voids is presented in Figure~\ref{fig:varchi:hotspot_delta-D} for different phase contrasts. As expected, the probability to find hard and soft phase around voids increases with the phase contrast, as the hard phase restricts the soft phase more for larger values of $\chi$. The same observation holds for the influence of the voids, confirming a weak dependence on the microstructure and other voids at low phase contrast and a strong dependence at high phase contrast.

\subsection{Combined effect}

\begin{figure}[htp]
  \centering
  \includegraphics[width=0.5\textwidth]{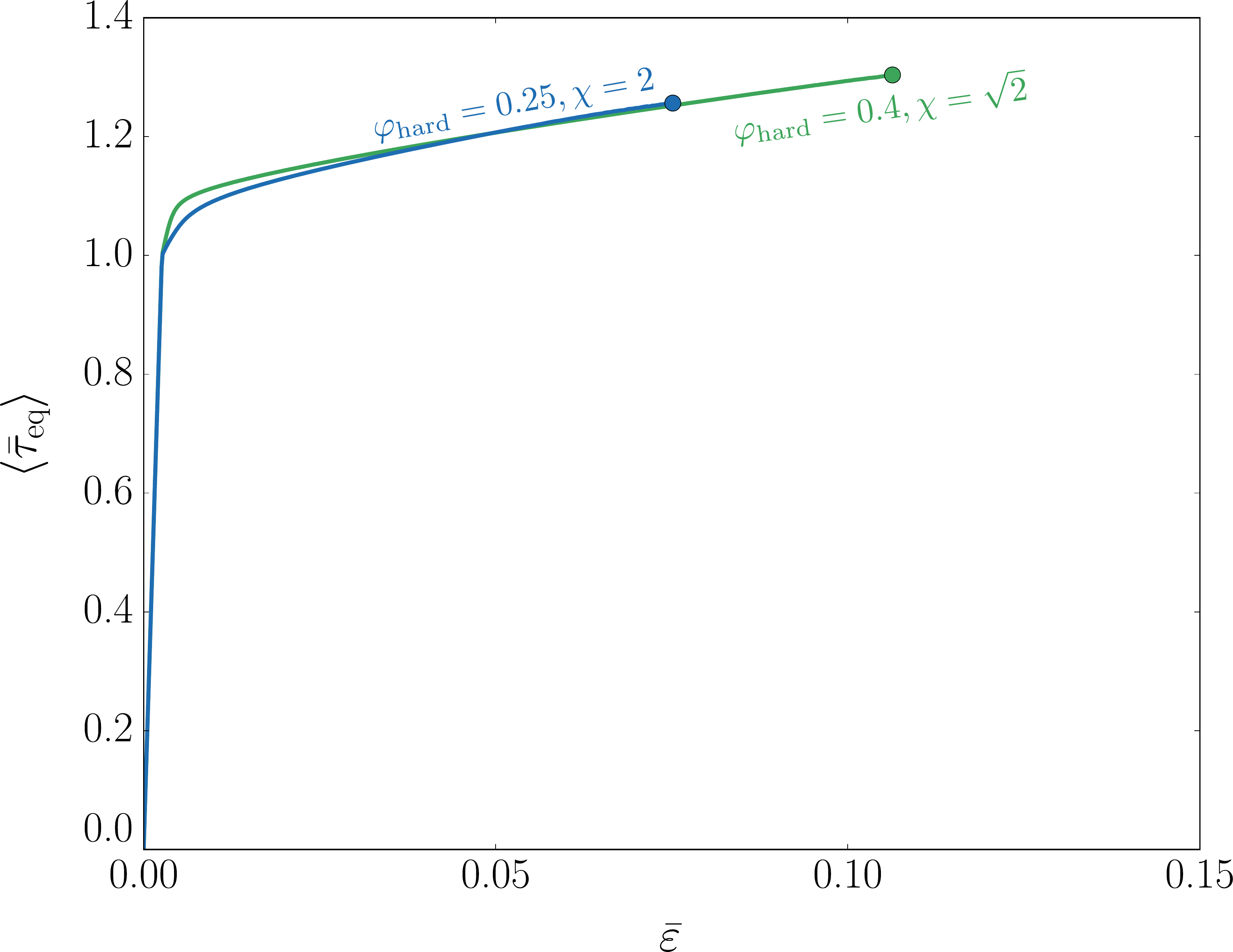}
  \caption{Combined influence of the hard phase volume fraction $\varphi_\mathrm{hard}$ and the phase contrast $\chi$ on the macroscopic response. Only the best matching responses are shown.}
  \label{fig:var:macro}
\end{figure}

Since the hard phase volume fraction and the contrast between the phases influence the macroscopic response in a similar way, one wonders if the ductility of the reference case can be enhanced by increasing the hard phase volume fraction while decreasing the phase contrast or vice versa. To this end, the volume fraction is varied in the range $\varphi_\mathrm{hard} = [0.1, 0.25, 0.4]$, and for each case the phase contrast is varied in the range $\chi = [\sqrt{2}, 2, 4]$. From all of the homogenized responses one parameter set is identified which results in a similar macroscopic hardening as for the reference case. The combination $\varphi_\mathrm{hard} = 0.4$ and $\chi = \sqrt{2}$ is found to have a particularly close match to $\varphi_\mathrm{hard} = 0.25$ and $\chi = 2$. These two responses are shown in Figure~\ref{fig:var:macro}. Obviously, in terms of both strength and ductility it is preferred to increase the hard phase volume fraction while decreasing the phase contrast. Note that the difference is quite significant: the fracture strain is increased by $0.03$. This lines up with earlier predictions by De Geus et al.~\citep{DeGeus2015a} that were made based on an indicator for fracture initiation only (that had no coupling to the mechanics whatsoever). Note that in reality the hard phase volume fraction and its hardness may not be varied independently without modifying the chemical composition of the material. For example for dual-phase steel, the hardness of the martensite is not independent of its volume fraction when the amount of carbon in the steel is kept constant (see e.g.\ Ref.~\citep{Choi2009} and references therein).

\section{Influence of stress state}
\label{sec:triax}

\subsection{Damage model and applied deformation}

While for pure shear the damage was concentrated in the soft phase, for different stress states, in particular those characterized by a large hydrostatic tensile stress, damage is expected to develop also in the hard phase \citep{DeGeus2015b}. The model is therefore modified in two ways. First, the hard phase is modeled quasi-brittle instead of infinitely strong by introducing a failure criterion for it as well. Second, different stress states are considered by adding a volumetric macroscopic strain to the pure shear deformation according to different ratios of macroscopic volumetric and deviatoric deformation. All other model ingredients are the same as above, whereby the same reference case is used, with the hard phase volume fraction $\varphi_\mathrm{hard} = 0.25$ and the phase contrast $\chi = 2$.

Damage and fracture in the hard phase are modeled using the damage indicator in Eq.~\eqref{eq:model:D}. The parameters
\begin{equation}
  A^\mathrm{hard} = 0.87 \quad
  B^\mathrm{hard} = 5.6  \quad
  \varepsilon_\mathrm{pc}^\mathrm{hard} = 0
\end{equation}
ensure a rapid decrease of the critical strain with increasing stress triaxiality. Compared to the soft phase, it is much lower for high stress triaxialities. The hard phase can therefore be thought of as relatively brittle compared to the soft phase \citep{Vajragupta2012,DeGeus2015b}.

Different stress states are applied by adding different amounts of volumetric deformation to the applied pure shear. Still assuming plane strain, the following logarithmic strain is prescribed
\begin{equation}
  \bar{\bm{\varepsilon}} =
  \tfrac{\sqrt{3}}{2} \, \bar{\varepsilon}_\mathrm{v}
  \left(
    \vec{e}_\mathrm{x} \vec{e}_\mathrm{x} +
    \vec{e}_\mathrm{y} \vec{e}_\mathrm{y}
  \right)
  +
  \tfrac{\sqrt{3}}{2} \, \bar{\varepsilon}_\mathrm{d}
  \left(
    \vec{e}_\mathrm{x} \vec{e}_\mathrm{x} -
    \vec{e}_\mathrm{y} \vec{e}_\mathrm{y}
  \right)
\end{equation}
Whereby the equivalent deviatoric strain $\bar{\varepsilon}_\mathrm{d}$ is applied as before. The value of the equivalent volumetric strain, $\bar{\varepsilon}_\mathrm{v}$, is determined such that a certain macroscopic stress triaxiality $\bar{\eta}$ is approximated. Note that $\bar{\eta} = \bar{\tau}_\mathrm{m} / \bar{\tau}_\mathrm{eq}$, i.e.\ the ratio of the macroscopic mean Kirchhoff stress and the macroscopic equivalent Kirchhoff stress. Since the constitutive response depends on the microstructure, it can be obtained only by homogenization. This is avoided by (i) assuming that, since the microstructure is elastically homogeneous and the void volume fraction remains small, the hydrostatic stress state may be approximated by $\bar{\tau}_\mathrm{m} = \sqrt{3} K \bar{\varepsilon}_\mathrm{v}$ (with $K$ the bulk modulus of both phases); (ii) approximating the equivalent stress by the yield stress of the soft phase, $\tau_\mathrm{y0}^\mathrm{soft}$. These approximations are (i) valid only up to void nucleation and (ii) acceptable as the observed hardening is much lower that $\tau_\mathrm{y0}^\mathrm{soft}$, see e.g.\ Figure~\ref{fig:config013:macroscopic}. For a target triaxiality $\bar{\eta}$ we thus apply a constant value of $\bar{\varepsilon}_\mathrm{v} = \tau_\mathrm{y0}^\mathrm{soft} / (\sqrt{3} K)$. Note that during the first increments, wherein the microstructure responds elasticity, $\bar{\varepsilon}_\mathrm{v}$ is proportional to $\bar{\varepsilon}_\mathrm{d}$ and thus lower than the specified value. Also note that it was found that the ensemble averaged stress triaxiality remains close to the prescribed value also after void nucleation. For the triaxialities that are considered below, the actual value $\langle \bar{\eta} \rangle$ is approximately $10\%$ lower than the target value at the onset of instability. This observation can directly be linked to the neglected hardening.

\subsection{Results}

\begin{figure}[htp]
  \centering
  \includegraphics[width=.5\textwidth]{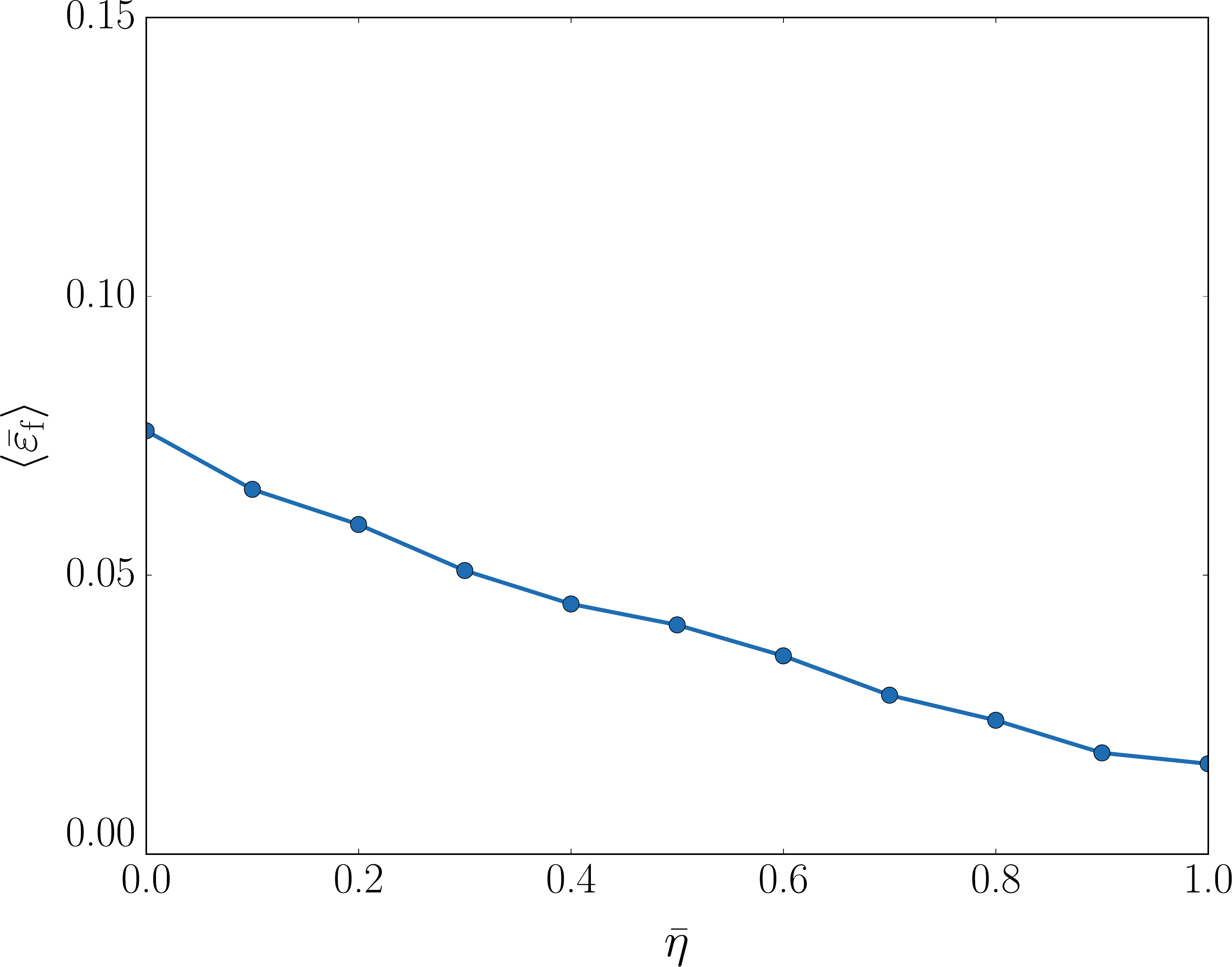}
  \caption{Ensemble averaged macroscopic fraction strain $\langle \bar{\varepsilon}_\mathrm{f} \rangle$ as a function of the applied (target) macroscopic stress triaxiality $\bar{\eta}$.}
  \label{fig:macroscopic_vartriax}
\end{figure}

The ensemble averaged macroscopic fracture initiation strain $\langle \bar{\varepsilon}_\mathrm{f} \rangle$ is plotted in Figure~\ref{fig:macroscopic_vartriax} for the different applied (target) macroscopic triaxialities $\bar{\eta}$. Eleven triaxialities are considered, evenly spaced in the range $\bar{\eta} = [0,1]$. It is observed that the fracture strain decreases as a function of $\bar{\eta}$. This can be understood from the observation that the local stress triaxialities are elevated in both phases, to accommodate the additional volumetric strain. This accelerates both damage mechanisms through Eq.~\eqref{eq:model:D}.

\begin{figure}[htp]
  \centering
  \includegraphics[width=1.\textwidth]{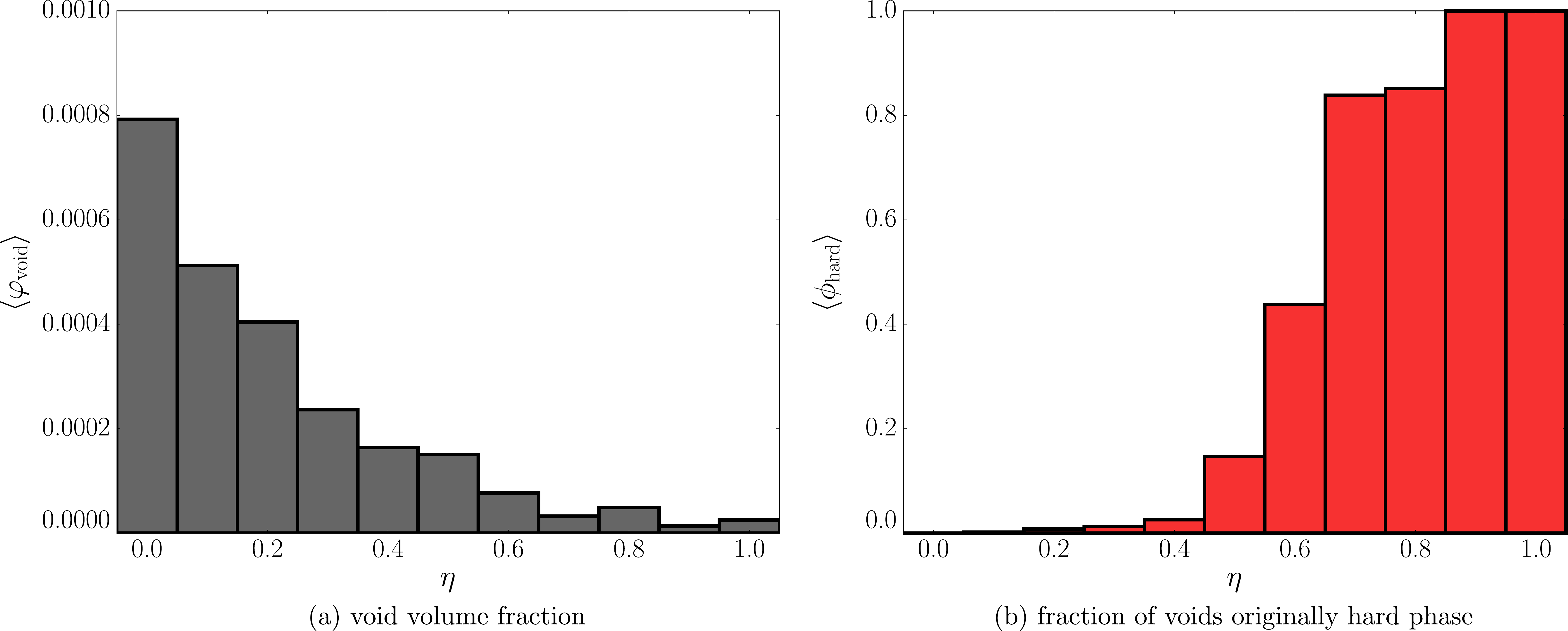}
  \caption{(a) Ensemble averaged void volume fraction $\langle \varphi_\mathrm{void} \rangle$, (b) fraction of voids nucleated in the hard phase $\langle \phi_\mathrm{hard} \rangle$; both as a function of the applied (target) macroscopic stress triaxiality $\bar{\eta}$.}
  \label{fig:voids_vartriax}
\end{figure}

The macroscopically observed apparent brittleness at elevated triaxialities is further inspected on the microscopic scale. The void volume fraction $\langle \varphi_\mathrm{void} \rangle$ at failure is shown in Figure~\ref{fig:voids_vartriax}(a) for the different applied macroscopic triaxialities. Indeed, instability occurs at lower void volume fractions at high triaxialities. In that regime, little dissipation by void nucleation is observed, corresponding to brittle fracture.

To understand the fracture mechanism at high triaxialities, the objective is to identify which of the two phases dominates. At the onset of instability this is done by determining the original phase of each void. The fraction of voids that have nucleated in the hard phase, $\langle \phi_\mathrm{hard} \rangle$, is shown in Figure~\ref{fig:voids_vartriax}(b) for the different values of applied macroscopic triaxiality. Note that the fraction of voids that have nucleated in the soft phase is simply $\langle \phi_\mathrm{soft} \rangle = 1 - \langle \phi_\mathrm{hard} \rangle$.

Figure~\ref{fig:voids_vartriax}(b) shows that at low triaxiality all voids nucleate in the soft phase. Upon increasing the triaxiality up to $\bar{\eta} = 0.6$ still the majority of voids are nucleated in the soft phase. At higher triaxialities ($\bar{\eta} > 0.6$) the majority of voids nucleate in the hard phase. At $\bar{\eta} = 1$ all voids have nucleated in the hard phase. It can be concluded that, up to the onset of instability, failure of the soft and the hard phase are in competition and the outcome of this competition depends on the stress triaxiality. Note also that it follows from Figure~\ref{fig:voids_vartriax}(b) that the assumption of the previous sections, that the hard phase does not fail, can be made without the loss of generality.

\begin{figure}[htp]
  \centering
  \includegraphics[width=.558\textwidth]{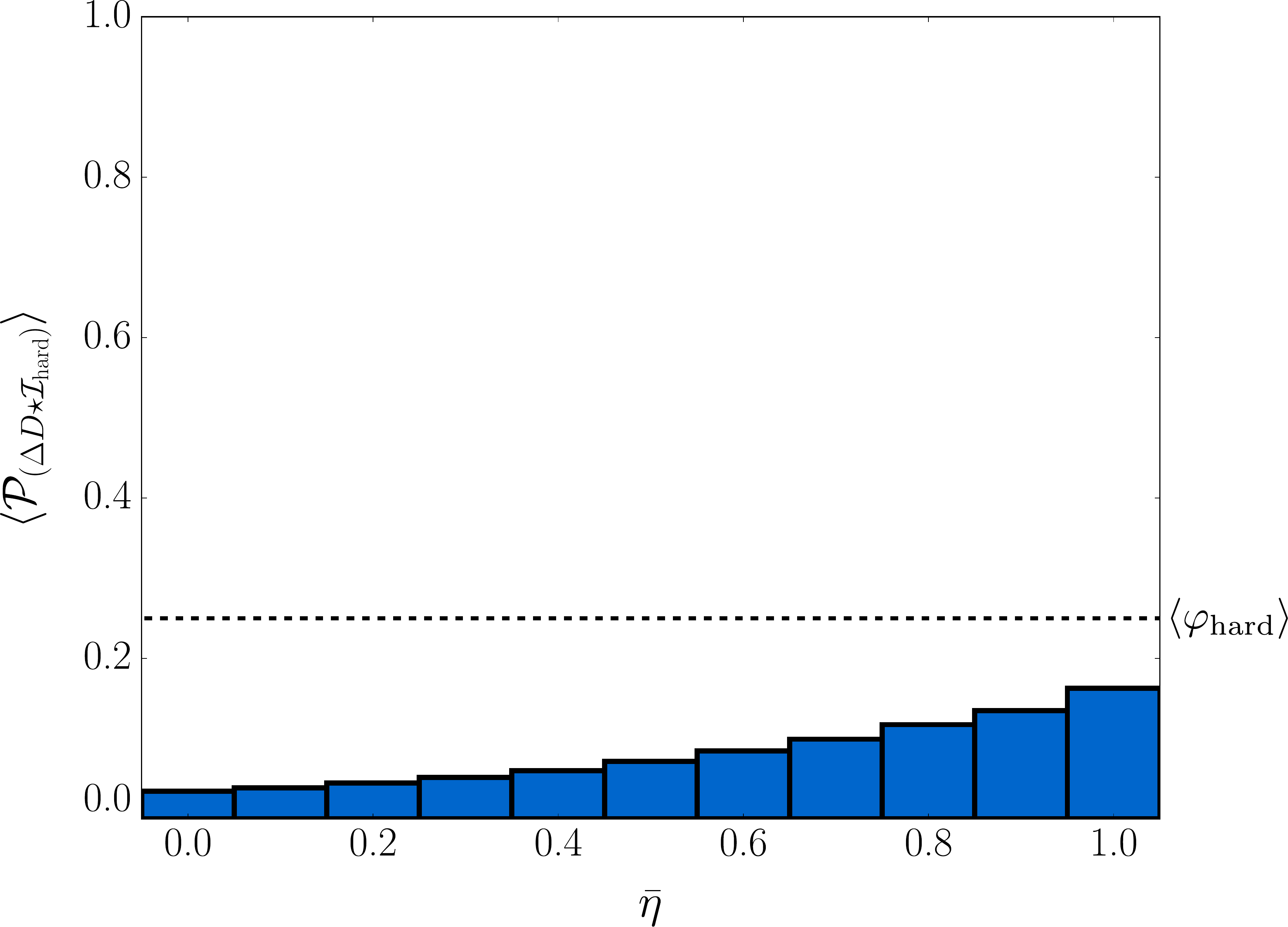}
  \caption{Probability of hard (and soft) ($\mathcal{P}_{(\Delta D \star \mathcal{I}_\mathrm{hard})}$) in cells with high incremental damage $\Delta D$, as a function of the applied (target) macroscopic stress triaxiality $\bar{\eta}$. The hard phase volume fraction $ \langle \varphi_\mathrm{hard} \rangle$ is indicated using a dashed line.}
  \label{fig:voids_vartriax_deltaD}
\end{figure}

To understand also the damage propagation beyond the onset of instability the damage activity, $\Delta D$, is again employed. The probability of hard phase is computed, but now only for the cell with greatest damage activity (i.e.\ Eq.~\eqref{eq:cor:P} at $\Delta \vec{x} = \vec{0}$). To interpret the result, in Figure~\ref{fig:voids_vartriax_deltaD}, $\mathcal{P}_{(\Delta D \star \mathcal{I}_\mathrm{hard})}$ once again should be compared to the hard phase volume fraction, $\langle \varphi_\mathrm{hard} \rangle$, shown using a dashed line. The observed $\mathcal{P}_{(\Delta D \star \mathcal{I}_\mathrm{hard})} < \langle \varphi_\mathrm{hard} \rangle$ for all triaxialities indicates an elevated probability of soft phase. It thus follows that, although all voids have nucleated in the hard phase, the damage activity is particularly high in the soft phase. The model thus predicts damage thus to propagate predominantly through the soft phase beyond the onset of instability.

\section{Discussion}
\label{sec:discusion}

\subsection{Model limitations}

Given the very idealized character of our model, it is in order that we discuss its limitations. In earlier publications we have found that the influence of the square cells is limited at the considered scale of observation (of individual cells) \citep{DeGeus2015}, but also that assuming isotropic plasticity is reasonable since the mechanics are dominated by the presence of a second phase with a very different yield stress \citep{DeGeus2016d}. By comparing with simulations of three-dimensional microstructures it was found that the influence of the phase distribution on damage can well be identified  using the two-dimensional model, but also that the value of the damage is over-predicted by the two-dimensional model \citep{DeGeus2016a}.

\begin{figure}[htp]
  \centering
  \includegraphics[width=1.\textwidth]{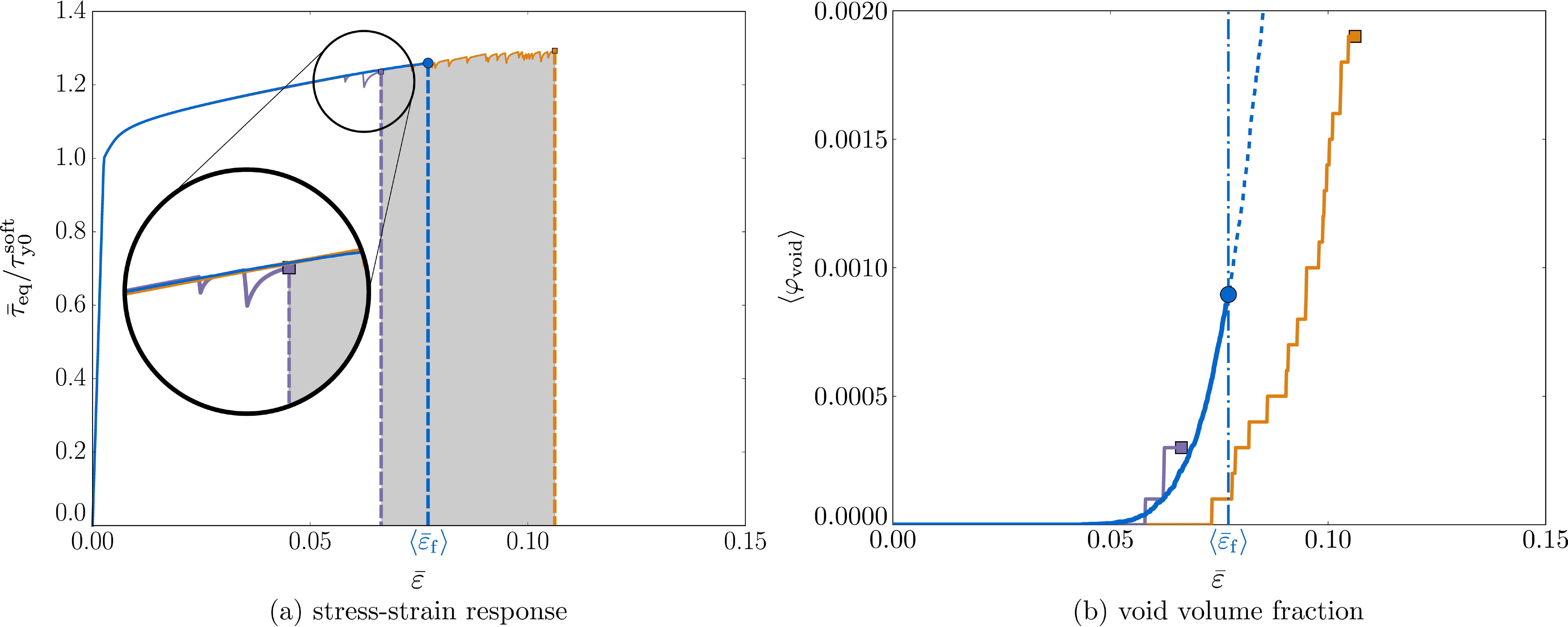}
  \caption{Macroscopic (a) stress--strain response and (b) void volume fraction, for an ensemble of $400$ microstructures comprising $50 \times 50$ cells with the reference parameters ($\varphi_\mathrm{hard} = 0.25$ and $\chi = 2$). For both, the homogenized response is shown in blue, the minimum of all microstructures in purple and the maximum in orange.}
  \label{fig:verify}
\end{figure}

In the context of the present contribution, the following precautions have been taken to present results that are nearly independent of the assumed periodicity: (i) the analysis is limited to the onset of global localization, (ii) a large volume element and (iii) a large statistical ensemble are used. To verify that the results are insensitive to the assumed periodicity, the full analysis is repeated for the reference case in Section~\ref{sec:reference}, using an ensemble of $400$ smaller volume elements, each with $50 \times 50$, instead of $100 \times 100$, cells. Note that the total number of cells is thus the same. All of the above observations are confirmed. Quantitative differences are found to be related to the sizes of the volume elements and the different stochastic properties of the ensembles, rather than to the boundary conditions. In particular, the scatter in the mechanical response, void volume fraction, and failure strain is somewhat larger due to the smaller volume element used and the larger ensemble. This is illustrated in Figure~\ref{fig:verify} through (a) the macroscopic stress--strain response and (b) the void volume fraction (cf.\ Figures~\ref{fig:config013:macroscopic} and \ref{fig:config013:voids} respectively). Based on these results we conclude that the trend is identical for both ensembles, while quantitatively the homogenized fracture strain is predicted a factor $1.03$ higher and the void volume fraction a factor $1.2$ higher using the $50 \times 50$ volume elements. In terms of the average phase and void distribution around incremental damage (cf.\ Figure~\ref{fig:config013:hotspot_delta-D}) the results match within $4\%$ accuracy; they are therefore not included here for brevity.

\subsection{Comparison with observations in the literature}

In accordance with the above results, \citet{Alaie2015} have found that voids nucleate often in between martensite islands (see also \citep{Uthaisangsuk2011,Avramovic-Cingara2009} and references in \citep{DeGeus2015a}). Upon macroscopic localization voids start to link up via ferrite (see also \citep{Zhou2015,Vajragupta2012,Avramovic-Cingara2009}). \citet{Avramovic-Cingara2009} thereby experimentally observe that damage propagation occurs by linking of voids along the shear direction. \citet{LLorca2004} have numerically found similar regions where the voids are nucleated: in the soft phase in-between hard phase particles that are closely separated along the tensile axis. They found that void growth only occurred in such features, up to the maximal stress. Our strongly idealized model thus reconciles observations that were made with much more involved models that consider both the mechanics as well as the morphology in much more detail. Because of the idealization, the present study adds a transparent analysis, in particular of the influence of the spatial distribution of the phases, featuring systematic variations. Thereby, the failure mechanisms are objectively identified using the statistical comparison of many different realizations.

In terms of macroscopic quantities, the exponential increase of the void volume fraction with the applied strain has also been observed experimentally (e.g.\ \citep{Avramovic-Cingara2009,Requena2013}). The same holds for the decrease in fracture strain with increasing triaxiality (e.g.\ \citep{Samei2016,Requena2013} and references in \citep{DeGeus2015b}). Quantitatively, our prediction of the values of the critical void volume fraction and fracture strain are small compared to experimentally observed values for comparable materials \citep{Bareggi2012,Avramovic-Cingara2009,Requena2013}. However, several differences in our model, of which the model parameters are the most important ones, may explain this mismatch. This is not further pursued here as the present study does not aim for quantitative predictions.

With respect to the influence of the hard phase volume fraction, \citet{Lai2015} reported that damage accumulation accelerates with increasing hard phase volume fraction (see also \citep{Uthaisangsuk2011}), which is in line with the present results. Circumstantial evidence supporting this can also be found in the work by \citet{Boselli2001}, who observed that fracture is accelerated in regions where the hard phase is closely separated. Besides the volume fraction also the statistical properties of the microstructure have been observed to affect the fracture mechanisms \citep{He1984,Kim1981}. This is been completely unexplored in the paper, and invites further research.

\clearpage
\section{Concluding remarks}
\label{sec:conclusion}

This paper has identified the mechanisms that lead from void nucleation to fracture of a two-phase material. This has been accomplished using an idealized model that facilitated a transparent statistical analysis. The following conclusions can be formulated.
\begin{itemize}
  \item After reaching an initial threshold, voids nucleate at all stages of deformation. Up to the point of instability the nucleation of voids remains dominant, dramatic damage propagation from a single void has not been observed.
  \item Damage initiation is observed in the soft phase at low stress triaxialities and in the hard phase at high triaxialities. It is governed by distinct local microstructural features. These `damage hot-spots' consist of a band of hard phase aligned with the tensile direction interrupted by bands of the soft phase aligned with the directions of maximum shear.
  \item Damage propagation from early nucleated voids is controlled on a longer length-scale than their initiation. Localization is triggered in a critical configuration wherein several voids are aligned with the directions of shear, with a percolation path linking them through the soft phase. This was shown by computing the correlation between damage propagation and voids.
  \item The moment of instability strongly depends how several initiation `hot-spots' are aligned with respect to each other. This results in a large scatter in fracture strains of different microstructures.
  \item With increasing hard phase volume fraction or phase contrast the relative influence of the microstructure as well as of the nucleated voids on the localization process is increased. In both cases, void nucleation is accelerated.
  \item Among the different combinations of volume fractions and phase contrasts which result in a similar hardening behavior, those with a high volume fraction and a low phase contrast reveal the largest fracture strain.
\end{itemize}
%

\section*{Acknowledgments}

This research was carried out under project number M22.2.11424 in the framework of the research program of the Materials innovation institute M2i (\href{http://www.m2i.nl}{www.m2i.nl}).

\clearpage
\section*{Nomenclature}

\subsection*{Notation}

\begin{tabular}{p{25mm}lll}
  $\bm{A}$
  & second order tensor
  \\
  $a$
  & scalar
  \\
  $\dot{a}$
  & rate
  \\
  $\Delta a$
  & time increment: $a(t + \Delta t) - a(t)$
  \\
  $\langle a \rangle$
  & ensemble average
  \\
  $\bar{a}$
  & volume average
  \\
\end{tabular}

\subsection*{Symbols}

\begin{tabular}{p{25mm}lll}
  $N$
  & number of cells in one volume element
  \\
  $\varphi$
  & volume fraction (e.g.\ $\varphi_\mathrm{hard}$ is the hard phase volume fraction)
  \\
  $\phi$
  & fraction
  \\
  $D (\vec{x})$
  & damage indicator
  \\
  $\mathcal{I} (\vec{x})$
  & phase (or void) indicator
  \\
  $\mathcal{P} (\vec{x})$
  & probability of a certain phase (or void)
  \\
  $\vec{e}_\mathrm{x}, \vec{e}_\mathrm{y}$
  & Cartesian basis vector (in $x$- and $y$-direction)
  \\
  $\vec{x}$
  & position vector
  \\
  $\bar{\bm{\varepsilon}}$
  & macroscopic logarithmic strain tensor
  \\
  $\bar{\varepsilon}$
  & macroscopic effective logarithmic strain
  \\
  $\bar{\varepsilon}_\mathrm{v}$
  & macroscopic volumetric effective logarithmic strain
  \\
  $\bar{\varepsilon}_\mathrm{d}$
  & macroscopic deviatoric effective logarithmic strain
  \\
  $\bar{\varepsilon}_\mathrm{f}$
  & macroscopic equivalent logarithmic strain at which fracture initiation is predicted
  \\
  $\bar{\tau}_\mathrm{f}$
  & macroscopic equivalent Kirchhoff stress at which fracture initiation is predicted
  \\
  $\bm{\varepsilon}_\mathrm{e} (\vec{x})$
  & logarithmic elastic strain tensor
  \\
  $\varepsilon_\mathrm{p} (\vec{x})$
  & effective plastic strain
  \\
  $\Phi (\vec{x})$
  & yield function
  \\
  $\bm{\tau} (\vec{x})$
  & Kirchhoff stress tensor
  \\
  $\tau_\mathrm{eq} (\vec{x})$
  & equivalent Kirchhoff stress
  \\
  $\tau_\mathrm{m} (\vec{x})$
  & hydrostatic Kirchhoff stress
  \\
  $\eta (\vec{x}) = \tau_\mathrm{m} / \tau_\mathrm{eq}$
  & stress triaxiality
  \\
\end{tabular}

\subsection*{Material constants}

\begin{tabular}{p{25mm}lll}
  $E$
  & Young's modulus
  \\
  $\nu$
  & Poisson's ratio
  \\
  $K$
  & bulk modulus
  \\
  $\tau_\mathrm{y0}$
  & initial Kirchhoff yield stress
  \\
  $H$
  & hardening modulus
  \\
  $\chi = \tau_\mathrm{y}^\mathrm{hard} / \tau_\mathrm{y}^\mathrm{soft}$
  & phase contrast: ratio of the (current) yield stress of the hard and of the soft phase
  \\
  $\varepsilon_\mathrm{c}$
  & critical effective plastic strain
  \\
  $A$, $B$
  & dependence of the critical strain $\varepsilon_\mathrm{c}$ on the stress triaxiality dependency $\eta$
  \\
  $\varepsilon_\mathrm{pc}$
  & critical effective plastic strain at infinitely high stress triaxiality
  \\
\end{tabular}

\clearpage
\scriptsize
\bibliography{library}

\end{document}